\documentclass[journal]{IEEEtran}
\usepackage{multicol}
\usepackage{graphicx}
\usepackage{latexsym}
\graphicspath{{Figures/}}
\usepackage{acronym}
\usepackage{amsmath}
\usepackage{color, colortbl}
\usepackage{subfigure}
\usepackage[english]{babel}
\usepackage{blindtext}
\usepackage{algorithm} 
\usepackage{algorithmic} 
\usepackage{multirow}
\usepackage{color}
\usepackage{subfigure}
\usepackage{amsfonts}
\usepackage{balance}
\usepackage{caption}
\usepackage{gensymb}

\begin{document}

\newacro{3GPP}{third generation partnership project}
\newacro{4G}{4{th} generation}
\newacro{5G}{5{th} generation}

\newacro{ADC}{analogue-to-digital conversion}
\newacro{AED}{accumulated euclidean distance}
\newacro{ASE}{amplified spontaneous emission}
\newacro{ASIC}{application specific integrated circuit}
\newacro{AWG}{arbitrary waveform generator}
\newacro{AWGN}{additive white Gaussian noise}
\newacro{A/D}{analog-to-digital}

\newacro{B2B}{back-to-back}
\newacro{BCF}{bandwidth compression factor}
\newacro{BCJR}{Bahl-Cocke-Jelinek-Raviv}
\newacro{BDM}{bit division multiplexing}
\newacro{BED}{block efficient detector}
\newacro{BER}{bit error rate}
\newacro{Block-SEFDM}{block-spectrally efficient frequency division multiplexing}
\newacro{BLER}{block error rate}
\newacro{BPSK}{binary phase shift keying}
\newacro{BS}{base station}
\newacro{BSS}{best solution selector}
\newacro{BU}{butterfly unit}

\newacro{CapEx}{capital expenditure}
\newacro{CA}{carrier aggregation}
\newacro{CBS}{central base station}
\newacro{CC}{component carriers}
\newacro{CCDF}{complementary cumulative distribution function}
\newacro{CCs}{component carriers}
\newacro{CD}{chromatic dispersion}
\newacro{CDF}{cumulative distribution function}
\newacro{CDI}{channel distortion information}
\newacro{CDMA}{code division multiple access}
\newacro{CI}{constructive interference}
\newacro{CIR}{carrier-to-interference power ratio}
\newacro{CMOS}{complementary metal-oxide-semiconductor}
\newacro{CNN}{convolutional neural network}
\newacro{CoMP}{coordinated multiple point}
\newacro{CO-SEFDM}{coherent optical-SEFDM}
\newacro{CP}{cyclic prefix}
\newacro{CPE}{common phase error}
\newacro{CRVD}{conventional real valued decomposition}
\newacro{CR}{cognitive radio}
\newacro{CRC}{cyclic redundancy check}
\newacro{CS}{central station}
\newacro{CSI}{channel state information}
\newacro{CSPR}{carrier to signal power ratio}
\newacro{CWT}{continuous wavelet transform}
\newacro{C-RAN}{cloud-radio access networks}

\newacro{DAC}{digital-to-analogue conversion}
\newacro{DBP}{digital backward propagation}
\newacro{DC}{direct current}
\newacro{DCT}{discrete cosine transform}
\newacro{DDC}{digital down-conversion}
\newacro{DDO-OFDM}{directed detection optical-OFDM}
\newacro{DDO-OFDM}{direct detection optical-OFDM}
\newacro{DDO-SEFDM}{directed detection optical-SEFDM}
\newacro{DFB}{distributed feedback}
\newacro{DFDMA}{distributed FDMA}
\newacro{DFT}{discrete Fourier transform}
\newacro{DFrFT}{discrete fractional Fourier transform}
\newacro{DMA}{direct memory access}
\newacro{DMRS}{demodulation reference signal}
\newacro{DOFDM}{dense orthogonal frequency division multiplexing}
\newacro{DP}{dual polarization}
\newacro{DPC}{dirty paper coding}
\newacro{DSB}{double sideband}
\newacro{DSL}{digital subscriber line}
\newacro{DSP}{digital signal processors}
\newacro{DVB}{digital video broadcast}
\newacro{DWT}{discrete wavelet transform}
\newacro{D/A}{digital-to-analog}

\newacro{ECC}{error correcting codes}
\newacro{ECL}{external-cavity laser}
\newacro{EDFA}{erbium doped fiber amplifier}
\newacro{EE}{energy efficiency}
\newacro{eMBB}{enhanced mobile broadband}
\newacro{eNB-IoT}{enhanced NB-IoT}
\newacro{EPA}{extended pedestrian A}
\newacro{EVM}{error vector magnitude}

\newacro{Fast-OFDM}{fast-orthogonal frequency division multiplexing}
\newacro{FBMC}{filterbank based multicarrier }
\newacro{FCE}{full channel estimation}
\newacro{FD}{fixed detector}
\newacro{FDD}{frequency division duplexing}
\newacro{FDM}{frequency division multiplexing}
\newacro{FDMA}{frequency division multiple access}
\newacro{FE}{full expansion}
\newacro{FEC}{forward error correction}
\newacro{FEXT}{far-end crosstalk}
\newacro{FF}{flip-flop}
\newacro{FFT}{fast Fourier transform}
\newacro{FIFO}{first in first out}
\newacro{F-OFDM}{filtered-orthogonal frequency division multiplexing}
\newacro{FPGA}{field programmable gate array}
\newacro{FrFT}{fractional Fourier transform}
\newacro{FSD}{fixed sphere decoding}
\newacro{FSD-MNSF}{FSD-modified-non-sort-free}
\newacro{FSD-NSF}{FSD-non-sort-free}
\newacro{FSD-SF}{FSD-sort-free}
\newacro{FSK}{frequency shift keying}
\newacro{FTN}{faster than Nyquist}
\newacro{FTTB}{fiber to the building}
\newacro{FTTC}{fiber to the cabinet}
\newacro{FTTdp}{fiber to the distribution point}
\newacro{FTTH}{fiber to the home}

\newacro{GB}{guard band}
\newacro{GFDM}{generalized frequency division multiplexing}
\newacro{GPU}{graphics processing unit}
\newacro{GSM}{global system for mobile communication}
\newacro{GUI}{graphical user interface}

\newacro{HC-MCM}{high compaction multi-carrier communication}
\newacro{HPA}{high power amplifier}

\newacro{IC}{integrated circuit}
\newacro{ICI}{inter carrier interference}
\newacro{ID}{iterative detection}
\newacro{IDCT}{inverse discrete cosine transform}
\newacro{IDFT}{inverse discrete Fourier transform}
\newacro{IDFrFT}{inverse discrete fractional Fourier transform}
\newacro{ID-FSD}{iterative detection-FSD}
\newacro{ID-SD}{ID-sphere decoding}
\newacro{IF}{intermediate frequency}
\newacro{IFFT}{inverse fast Fourier transform}
\newacro{IFrFT}{inverse fractional Fourier transform}
\newacro{IMD}{intermodulation distortion}
\newacro{IoT}{internet of things}
\newacro{IOTA}{isotropic orthogonal transform algorithm}
\newacro{IP}{intellectual property}
\newacro{ISC}{interference self cancellation}
\newacro{ISI}{inter symbol interference}

\newacro{LDPC}{low density parity check}
\newacro{LFDMA}{localized FDMA}
\newacro{LLR}{log-likelihood ratio}
\newacro{LNA}{low noise amplifier}
\newacro{LO}{local oscillator}
\newacro{LOS}{line-of-sight}
\newacro{LPWAN}{low power wide area network}
\newacro{LS}{least square}
\newacro{LTE}{long term evolution}
\newacro{LTE-Advanced}{long term evolution-advanced}
\newacro{LUT}{look-up table}

\newacro{MA}{multiple access}
\newacro{MAC}{media access control}
\newacro{MASK}{m-ary amplitude shift keying}
\newacro{MCM}{multi-carrier modulation}
\newacro{MC-CDMA}{multi-carrier code division multiple access}
\newacro{MCS}{modulation and coding scheme}
\newacro{MF}{matched filter}
\newacro{MIMO}{multiple input multiple output}
\newacro{ML}{maximum likelihood}
\newacro{MLSD}{maximum likelihood sequence detection}
\newacro{MMF}{multi-mode fiber}
\newacro{MMSE}{minimum mean squared error}
\newacro{mMTC}{massive machine-type communication}
\newacro{MNSF}{modified-non-sort-free}
\newacro{MOFDM}{masked-OFDM}
\newacro{MRVD}{modified real valued decomposition}
\newacro{MS}{mobile station}
\newacro{MSE}{mean squared error}
\newacro{MTC}{machine-type communication}
\newacro{MUSA}{multi-user shared access}
\newacro{MU-MIMO}{multi-user multiple-input multiple-output}
\newacro{MZM}{Mach-Zehnder modulator}
\newacro{M2M}{machine to machine}

\newacro{NB-IoT}{narrowband IoT}
\newacro{NEXT}{near-end crosstalk}
\newacro{NG-IoT}{next generation IoT}
\newacro{NLOS}{non-line-of-sight}
\newacro{NN}{neural network}
\newacro{NOFDM}{non-orthogonal frequency division multiplexing}
\newacro{NOMA}{non-orthogonal multiple access}
\newacro{NoFDMA}{non-orthogonal frequency division multiple access}
\newacro{NP}{non-polynomial}
\newacro{NR}{new radio}
\newacro{NSF}{non-sort-free}
\newacro{NWDM}{Nyquist wavelength division multiplexing }
\newacro{Nyquist-SEFDM}{Nyquist-spectrally efficient frequency division multiplexing}

\newacro{OBM-OFDM}{orthogonal band multiplexed OFDM}
\newacro{OF}{optical filter}
\newacro{OFDM}{orthogonal frequency division multiplexing}
\newacro{OFDMA}{orthogonal frequency division multiple access}
\newacro{OMA}{orthogonal multiple access}
\newacro{OpEx}{operating expenditure}
\newacro{OQAM}{offset-QAM}
\newacro{OSNR}{optical signal-to-noise ratio}
\newacro{OSSB}{optical single sideband}
\newacro{OTA}{over-the-air}
\newacro{Ov-FDM}{Overlapped FDM}
\newacro{O-SEFDM}{optical-spectrally efficient frequency division multiplexing}
\newacro{O-FOFDM}{optical-fast orthogonal frequency division multiplexing}
\newacro{O-OFDM}{optical-orthogonal frequency division multiplexing}

\newacro{PA}{power amplifier}
\newacro{PAPR}{peak-to-average power ratio}
\newacro{PCE}{partial channel estimation}
\newacro{PD}{photodiode}
\newacro{PDF}{probability density function}
\newacro{PDP}{power delay profile}
\newacro{PDMA}{polarisation division multiple access}
\newacro{PDM-OFDM}{polarization-division multiplexing-OFDM}
\newacro{PDM-SEFDM}{polarization-division multiplexing-SEFDM}
\newacro{PDSCH}{physical downlink shared channel}
\newacro{PE}{processing element}
\newacro{PED}{partial Euclidean distance}
\newacro{PMD}{polarization mode dispersion}
\newacro{PON}{passive optical network}
\newacro{PPM}{parts per million}
\newacro{PRB}{physical resource block}
\newacro{PSD}{power spectral density}
\newacro{PU}{primary user}
\newacro{PXI}{PCI extensions for instrumentation}
\newacro{P/S}{parallel-to-serial}

\newacro{QAM}{quadrature amplitude modulation}
\newacro{QoS}{quality of service}
\newacro{QPSK}{quadrature phase-shift keying}

\newacro{RAUs}{remote antenna units}
\newacro{RBW}{resolution bandwidth}
\newacro{RF}{radio frequency}
\newacro{RMS}{root mean square}
\newacro{RoF}{radio-over-fiber}
\newacro{ROM}{read only memory}
\newacro{RRC}{root raised cosine}
\newacro{RSC}{recursive systematic convolutional}
\newacro{RTL}{register transfer level}
\newacro{RVD}{real valued decomposition}

\newacro{ScIR}{sub-carrier to interference ratio}
\newacro{SCMA}{sparse code multiple access}
\newacro{SC-FDMA}{single carrier frequency division multiple access}
\newacro{SC-SEFDMA}{single carrier spectrally efficient frequency division multiple access}
\newacro{SD}{sphere decoding}
\newacro{SDP}{semidefinite programming}
\newacro{SDR}{software defined radio}
\newacro{SE}{spectral efficiency}
\newacro{SEFDM}{spectrally efficient frequency division multiplexing}
\newacro{SEFDMA}{spectrally efficient frequency division multiple access} 
\newacro{SF}{sort-free}
\newacro{SIC}{successive interference cancellation}
\newacro{SiGe}{silicon-germanium}
\newacro{SINR}{signal-to-interference-plus-noise ratio}
\newacro{SISO}{single-input single-output}
\newacro{SMF}{single mode fiber}
\newacro{SNR}{signal-to-noise ratio}
\newacro{SP}{shortest-path}
\newacro{SRS}{sounding reference signal}
\newacro{SSB}{single-sideband}
\newacro{SSBI}{signal-signal beat interference}
\newacro{SSMF}{standard single mode fiber}
\newacro{STBC}{space time block coding}
\newacro{STO}{symbol timing offset}
\newacro{SU}{secondary user}
\newacro{SVD}{singular value decomposition}
\newacro{SVR}{singular value reconstruction}
\newacro{S/P}{serial-to-parallel}

\newacro{TDD}{time division duplexing}
\newacro{TDMA}{time division multiple access }
\newacro{TFP}{time frequency packing}
\newacro{THP}{Tomlinson-Harashima precoding}
\newacro{TOFDM}{truncated OFDM}
\newacro{TSVD}{truncated singular value decomposition}
\newacro{TSVD-FSD}{truncated singular value decomposition-fixed sphere decoding}

\newacro{UCR}{user compression ratio}
\newacro{UFMC}{universal-filtered multi-carrier}
\newacro{URLLC}{ultra-reliable and low-latency communication}
\newacro{USRP}{universal software radio peripheral}

\newacro{VDSL}{very-high-bit-rate digital subscriber line}
\newacro{VDSL2}{very-high-bit-rate digital subscriber line 2}
\newacro{VHDL}{very high speed integrated circuit hardware description language}
\newacro{VLC}{visible light communication}
\newacro{VLSI}{very large scale integration}
\newacro{VOA}{variable optical attenuator}
\newacro{VP}{vector perturbation}
\newacro{VSSB-OFDM}{virtual single-sideband OFDM}

\newacro{WAN}{wide area network}
\newacro{WCDMA}{wideband code division multiple access}
\newacro{WDM}{wavelength division multiplexing}
\newacro{WiFi}{wireless fidelity}
\newacro{WiGig}{Wireless Gigabit Alliance}
\newacro{WiMAX}{Worldwide interoperability for Microwave Access}
\newacro{WSS}{wavelength selective switch}

\newacro{ZF}{zero forcing}
\newacro{ZP}{zero padding}


\title{Design and Prototyping of Hybrid Analogue Digital Multiuser MIMO Beamforming for\\ Non-Orthogonal Signals }

\author{{Tongyang Xu,~\IEEEmembership{Member,~IEEE}, Christos Masouros,~\IEEEmembership{Senior Member,~IEEE} and Izzat Darwazeh,~\IEEEmembership{Senior Member,~IEEE}}
\thanks{
T. Xu, C. Masouros and I. Darwazeh are with the Department of Electronic and Electrical Engineering, University College London (UCL), London, WC1E 7JE, UK (e-mail: tongyang.xu.11@ucl.ac.uk, c.masouros@ucl.ac.uk, i.darwazeh@ucl.ac.uk).
}}

\maketitle

\begin{abstract}

To enable user diversity and multiplexing gains, a fully digital precoding multiple input multiple output (MIMO) architecture is typically applied. However, a large number of radio frequency (RF) chains make the system unrealistic to low-cost communications. Therefore, a practical three-stage hybrid analogue-digital precoding architecture, occupying fewer RF chains, is proposed aiming for a non-orthogonal IoT signal in low-cost multiuser MIMO systems. The non-orthogonal waveform can flexibly save spectral resources for massive devices connections or improve data rate without consuming extra spectral resources. The hybrid precoding is divided into three stages including analogue-domain, digital-domain and waveform-domain. A codebook based beam selection simplifies the analogue-domain beamforming via phase-only tuning. Digital-domain precoding can fine-tune the codebook shaped beam and resolve multiuser interference in terms of both signal amplitude and phase. In the end, the waveform-domain precoding manages the self-created inter carrier interference (ICI) of the non-orthogonal signal. This work designs over-the-air signal transmission experiments for fully digital and hybrid precoding systems on software defined radio (SDR) devices. Results reveal that waveform precoding accuracy can be enhanced by hybrid precoding. Compared to a transmitter with the same RF chain resources, hybrid precoding significantly outperforms fully digital precoding by up to 15.6 dB error vector magnitude (EVM) gain. A fully digital system with the same number of antennas clearly requires more RF chains and therefore is low power-, space- and cost- efficient. Therefore, the proposed three-stage hybrid precoding is a quite suitable solution to non-orthogonal IoT applications.

\end{abstract}


\section{Introduction}

\IEEEPARstart{W}{ith} the development of industry 4.0, IoT devices are becoming multi-functional but with increasing system complexity. There are three significant challenges in current IoT devices: to extend the signal transmission distance, more power has to be consumed; to aggregate massive devices, extra spectral or timing resources have to be occupied; to boost data rate, high-order modulation formats have to be applied. It is predicted that the world would have 55 billion connected devices by 2025, and to support such rapidly growing IoT services in wider signal coverage, massive device connections and increased data rate, new techniques have to be developed.

In the \ac{4G} \cite{Erik_book_4G} and \ac{5G} \cite{Erik_book_5G}, orthogonal \ac{IoT}, such as \ac{OFDM} based \ac{NB-IoT}, is no longer efficient for \ac{NG-IoT} requirements. Non-orthogonal signal waveforms, such as \ac{Fast-OFDM} \cite{FOFDM2002,Tongyang_NB_IoT_2018,Fast-OFDM-index_mod_CL2019}, \ac{SEFDM} \cite{SEFDM2003,Izzat_CSNDSP2018,SEFDM-index_mod_WCL2019,SEFDM_IoT_2018}, \ac{FBMC} \cite{NB_IoT_FBMC_2019}, Half-Sinc \cite{Tongyang_PIMRC2018_Half_sinc}, nonorthogonal \ac{FSK} \cite{Nonorthogonal_FSK_Luca_2018} and Hilbert-pair waveform \cite{xinyue_Hilbert_VTC2019}, bring benefits either in signal bandwidth saving or data rate improvement. The 4G/5G standards maintained the use of OFDM for NB-IoT as there are benefits of compatibility with the 4G/5G signal formats and the general benefits of multicarrier signals in simply and simultaneously correcting imperfect timing synchronization, \ac{LO} phase offset, sampling phase offset and other joint hardware/channel impairments. This work will focus on the enhancement of downlink NB-IoT, which employs the traditional multicarrier OFDM signal. The bandwidth compression SEFDM signal waveform is investigated in this work primarily for its spectral efficiency enhancement and energy efficiency improvement relative to OFDM as well as for its ease of implementation and backward compatibility. With careful waveform scheduling \cite{Tongyang_GLOBECOM2018}, power efficiency can be significantly enhanced leading to battery life extension. These advantages are apparent but at the cost of \ac{ICI}. For uplink channels, advanced signal detectors can be applied since signal processing is within base stations. In such cases, the uplink IoT devices capacity can be improved. For downlink channels, precoding has to be used to simplify each IoT device.

Although single antenna is preferred in IoT scenarios for the purpose of simplicity, future IoT applications would require services based on multiantenna multiuser techniques. Multiple input multiple output (MIMO) is now widely used to improve capacity and achieve diversity and multiplexing gains. There are a number of renowned MIMO testbeds such as Argos \cite{Argos_2012}, Ngara \cite{Ngara_2012}, TitanMIMO \cite{TitanMIMO_2015}, Lund University's MIMO \cite{Lund_MIMO_2014} and Bristol University's MIMO \cite{Bristol_MIMO_2016}. To achieve high diversity and multiplexing gains, pure digital precoding is normally used to pre-equalize inter-user interference in MIMO systems. The methodology is to modify amplitude and phase of symbols at baseband and then up convert the signal to \ac{RF} domain. In the end, antennas are connected to deliver the RF signal over the air. In order to achieve this, each antenna has to be controlled by one \ac{RF} chain, which is too expensive for multiantenna multiuser IoT systems.

Previous work has revealed the feasibility of using pure digital precoding to recover non-orthogonal SEFDM signals in the UCL multiuser MIMO-SEFDM testbed \cite{Tongyang_MU_MIMO_NB_IoT_2018}. However, six transmitter RF chains, associated with six antennas, have to be used to support two users. Analogue beamforming, such as the techniques used in IEEE 802.11ad and IEEE 802.15.3c, can greatly reduce hardware design complexity by using phase shifters. However, the analogue beamforming cannot achieve user multiplexing since only one beam can be generated, via phase shifters, per RF chain. This challenge can be solved by a hybrid analogue-digital precoding architecture \cite{hybrid_precoding_JSTSP2014, hybrid_precoding_CM2014, Hybrid_precoding_JSAC_2017, hybrid_precoding_TSP2017, hybrid_precoding_WCL2014, hybrid_precoding_CM2017}, which employs low-power and low-cost phase shifters to expand low-dimension MIMO systems to high-dimension MIMO systems to enable diversity and multiplexing gains. The strategy is to first apply phase shifters for phase-only analogue precoding and then employ the typical low-dimension baseband \ac{ZF} digital precoding to modify both signal amplitude and phase. The second-stage digital precoding is a fine-tune operation, which further mitigates the inter-user interference. The joint RF analogue and baseband digital precoding can greatly cut the hardware costs and achieve superior diversity and multiplexing gains. Existing MIMO hybrid precoding methods are designed for single carrier waveforms or the orthogonal multicarrier OFDM waveform. Since our target multicarrier SEFDM signal waveform breaks the orthogonality, precoding strategies have to be different.

This work designs a three-stage hybrid precoding architecture to enable a low-cost multiuser MIMO-SEFDM system. The first stage is analogue precoding, which changes the phase of each data stream leading to a spatially narrow beam. The second stage is digital precoding, which modifies both signal amplitude and phase to fine-tune the beam derived from the first precoding stage and mitigate inter-user interference. The third stage is waveform precoding, which pre-equalizes the interfered SEFDM signals. The three precoding stages are inter-related. The first stage shapes the beam using phase shifters, whose quality is crucial to relax the burden of the second stage fine-tuning. In addition, the first two precoding stages determine the last stage waveform precoding accuracy. Since a precoding matrix is multiplied in the third stage to pre-equalize the non-orthogonal signal, any deviations from the first two precoding stages would be amplified. This work will comprehensively investigate the joint three-stage precoding architecture and applies experiments to validate the performance.

The main contributions of this work are:

\begin{itemize}
\item{Novel three-stage precoding architecture for the hybrid analogue-digital multiuser MIMO-SEFDM system. The entire system can be optimized independently; } 

\item{Low-complexity analogue beam selection strategy with the assistance of staggered pilot symbols. The pilots can extract spatial channel information, which is used simultaneously for digital precoding and beam power measurement;}
 
\item{Experiment over-the-air prototyping of the three-stage hybrid analogue-digital system. In particular, extensive comparisons are provided via implementing two additional fully digital precoding multiuser MIMO systems; } 

\item{Enhanced hardware efficiency. Measured results reveal that the three-stage MIMO-SEFDM architecture outperforms fully digital MIMO-SEFDM with significant \ac{BER} and \ac{SE} gains occupying the same RF chain resources. To reach similar performance, the fully digital system has to use more RF chain resources resulting in reduced \ac{EE}. } 
\end{itemize}

\section{System Model}

\subsection{Waveform Model}

\begin{figure}
\begin{center}
\includegraphics[scale=0.48]{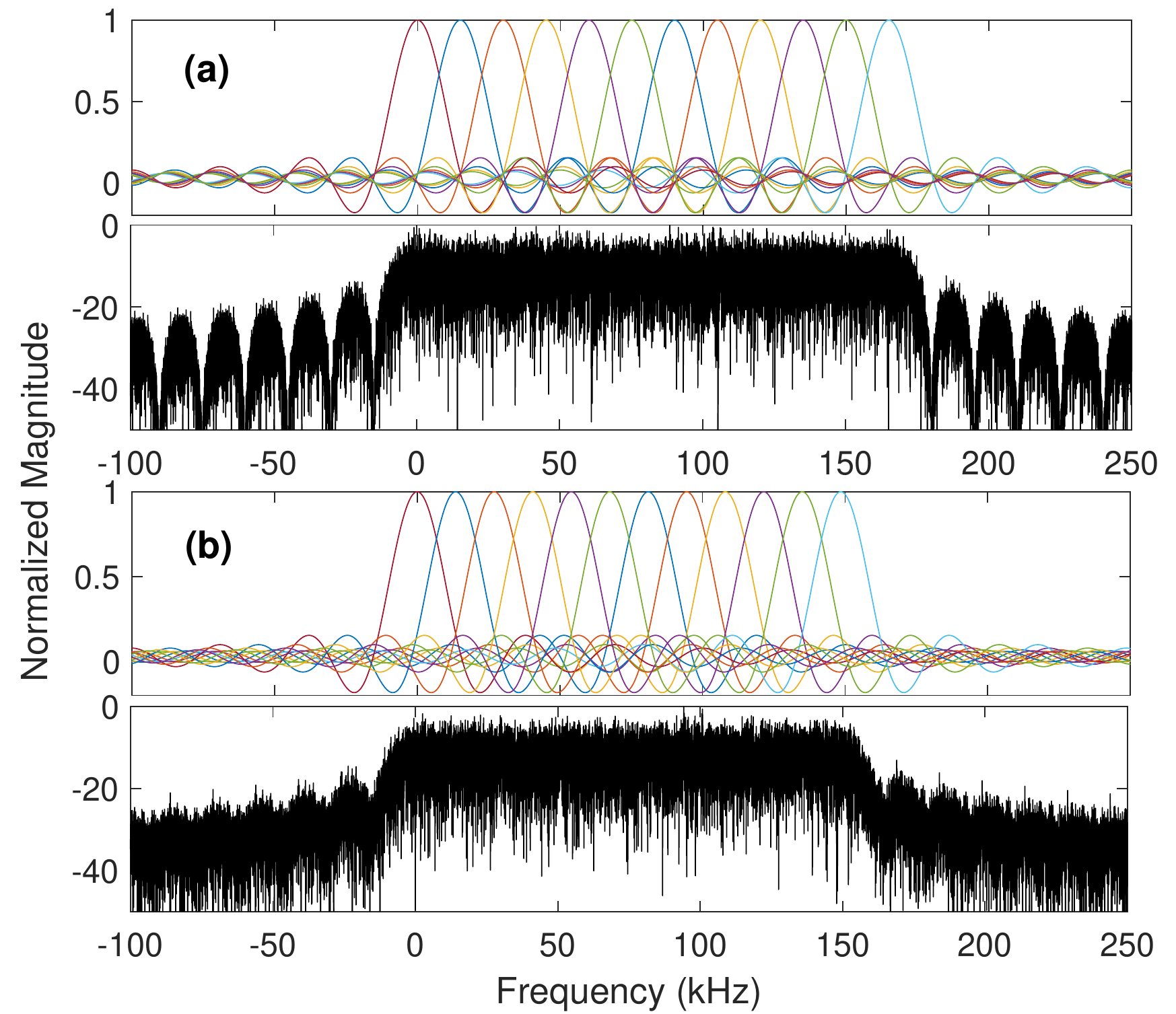}
\end{center}
\caption{{ Illustrative spectra for different multicarrier signals. (a) OFDM (12 sub-carriers, bandwidth is $B$). (b) SEFDM (12 sub-carriers, bandwidth compression factor $\alpha$=0.9, bandwidth is $\alpha\times{B}$).} }
\label{Fig:JIoT_spectral_response}
\end{figure}

The target IoT signal is the non-orthogonal multicarrier waveform \ac{SEFDM}, which packs sub-carriers closer than OFDM leading to compressed spectral bandwidth and therefore improved spectral efficiency, at the expense of self-created \ac{ICI} \cite{SafaCSNDSP2012}. {The sub-carrier packing strategy and frequency responses of OFDM and SEFDM signals are illustrated in Fig. \ref{Fig:JIoT_spectral_response}.}

\begin{figure*}[ht]
\begin{center}
\includegraphics[scale=0.46]{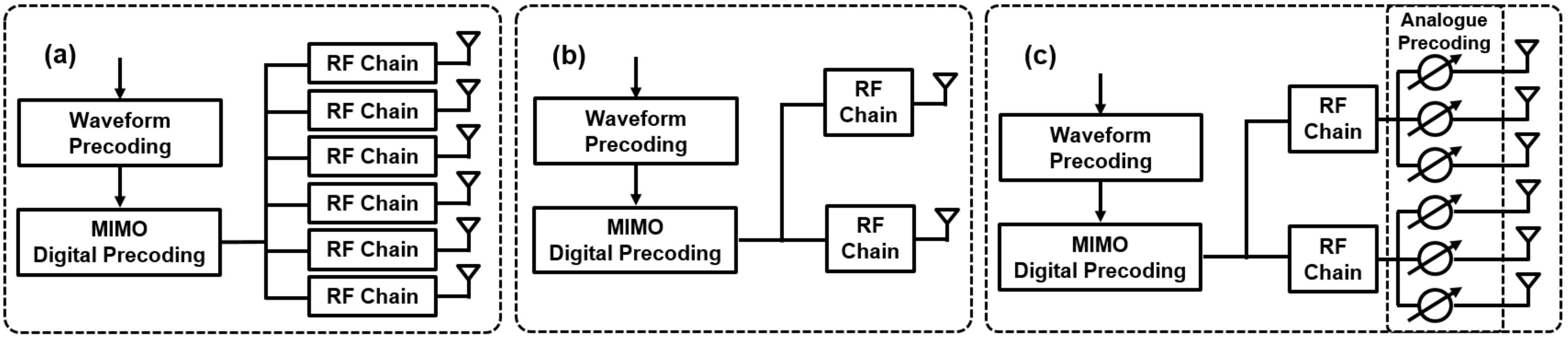}
\end{center}
\caption{Block diagrams of different precoding strategies. (a) Fully digital precoding (FDP-I) system with $N_{RF}$=6 RF chains and $N_{TX}$=6 transmitter antennas. (b) Fully digital precoding (FDP-II) system with $N_{RF}$=2 RF chains and $N_{TX}$=2 transmitter antennas. (c) Three-stage hybrid analogue-digital precoding (HP) system with $N_{RF}$=2 RF chains and $N_{TX}$=6 transmitter antennas.}
\label{Fig:Hybrid_beamforming_block_diagram}
\end{figure*}

The mathematical definition of a discrete $N$ sub-carrier SEFDM signal waveform $X_k$, in which $k$ is the time sample index ranging from 1 to $N$, is expressed as

\begin{equation}
X_k=\frac{1}{\sqrt{N}}\sum_{n=1}^{N}s_{n}\exp\left(\frac{j2{\pi}nk\alpha}N\right),\label{eq:SEFDM_discrete_signal}\end{equation}
where $\alpha=\Delta{f}\cdotp{T}$ is the bandwidth compression factor, which is the factor determines the sub-carrier packing characteristics. The signal is OFDM when $\alpha=1$ while $\alpha<1$ indicates the SEFDM signal. $\Delta{f}$ is the sub-carrier spacing and $T=N\cdotp{T_s}$ is the time period of one SEFDM symbol where $T_s$ is the time period of one sample. $s$ is an $N$-dimension vector consisting of QAM/PSK symbols, $n$ is the sub-carrier index ranging from 1 to $N$. {The matrix representation of the discrete SEFDM signal in \eqref{eq:SEFDM_discrete_signal} is expressed as 
\begin{equation}
X=\mathbf{F}S,\label{eq:Tx_signal_model}\end{equation}
where $S$ is an $N$-dimension vector consisting of $s_n$ and $\mathbf{F}$ is the $N\times{N}$ SEFDM modulation matrix, in which each element is defined as $\exp\left({j2{\pi}nk\alpha}/{N}\right)$. To clarify simply the principle of SEFDM signal waveform, only \ac{AWGN} is considered in this section. Therefore, the matrix format of a received SEFDM signal is defined as}
\begin{equation}
Y=\mathbf{F}S+Z,\label{eq:Rx_signal_model}\end{equation}
where $Z$ is an $N$-dimension vector of \ac{AWGN} samples.

At the receiver, the SEFDM signal of \eqref{eq:Rx_signal_model} is demodulated by multiplying a \ac{FrFT} matrix as
\begin{eqnarray}\label{eq:signal_rx_model}
R=\mathbf{F^{*}}(\mathbf{F}S+Z)=\mathbf{C}S+\mathbf{F^{*}}Z=\mathbf{C}S+Z_{\mathbf{F^{*}}},\end{eqnarray}
where $\mathbf{C}$ is the correlation matrix consisting of elements $c(m,n)$, which is defined by correlating two random $m^{th}$ and $n^{th}$ sub-carriers leading to an expression as 
\begin{equation} \label{eq:element_corr}
\begin{split}
  c_{m,n}&=\frac{1}{{N}}\sum_{k=1}^{N}e^{\frac{j2{\pi}mk\alpha}N}e^{-\frac{j2{\pi}nk\alpha}N}\\ &=\left\{
  \begin{array}{l l}
    1 & \quad \text{$,\ m=n$}\\
    \frac{1-e^{j2\pi\alpha(m-n)}}{N(1-e^{\frac{j2\pi\alpha(m-n)}{N}})} & \quad \text{$,\ m\neq{n}$}
  \end{array} \right\}. 
\end{split}
\end{equation}

The self-created ICI is given by the second term when $m\neq{n}$. It is clearly seen that the term equals zero for OFDM when $\alpha=1$ while it is non-zero when $\alpha{\neq}1$.

\subsection{Multiuser MIMO-SEFDM System Architectures}

We present three multiuser MIMO-SEFDM architectures in Fig. \ref{Fig:Hybrid_beamforming_block_diagram} that compares our fully digital precoding benchmarks and the proposed hybrid precoding. Fig. \ref{Fig:Hybrid_beamforming_block_diagram}(a) demonstrates our benchmark-I system, which is a fully digital precoding system with $N_{RF}$=6 RF chains and $N_{TX}$=6 transmitter antennas. This system architecture occupies more RF chain resources than the other two systems. This would bring reasonable performance but low hardware and energy efficiency. Fig. \ref{Fig:Hybrid_beamforming_block_diagram}(b) shows our benchmark-II system, which is also a fully digital precoding system but with $N_{RF}$=2 RF chains and $N_{TX}$=2 transmitter antennas. This architecture cuts the number of RF chains. Hardware efficiency is improved but performance would be affected. Fig. \ref{Fig:Hybrid_beamforming_block_diagram}(c) presents our proposed hybrid analogue-digital precoding system with $N_{RF}$=2 RF chains and $N_{TX}$=6 transmitter antennas, in which the analogue-domain is adjustable via low-cost phase shifters. This architecture maintains reasonable hardware efficiency and would be expected to achieve the performance similar to Fig. \ref{Fig:Hybrid_beamforming_block_diagram}(a).

\section{Principle of Three-Stage Hybrid Precoding}

The main idea of the three-stage hybrid precoding is to use jointly the Stage-I analogue precoding and the Stage-II digital precoding to enhance the accuracy of the Stage-III signal waveform precoding. System modelling in this work is based on single-antenna users. For the sake of practical evaluations, this work applies a sub-connected hybrid precoding architecture \cite{hybrid_precoding_CM2017}, which avoids connections crossing different RF chains and therefore simplifies the entire system design.

\subsection{Stage-I: MIMO Analogue Precoding} \label{subsec:MIMO_analogue_precoding}

In order to realize analogue precoding, multiple phase shifters are combined to an RF chain via power splitters and each phase shifter is connected to an omni-directional antenna. Existing work in \cite{hybrid_precoding_TSP2017, hybrid_precoding_WCL2014, hybrid_precoding_TWC2018,hybrid_precoding_JSAC2009} seek simplified analogue beamforming strategies and convert high-dimension hybrid analogue-digital precoding systems to equivalent low-dimension digital precoding systems. Work in \cite{hybrid_precoding_WSA2018, hybrid_precoding_ICASSP2019} designed circuits for a hybrid precoding analogue beamformer based on a predefined codebook. In this work, we follow the similar strategies by introducing a predefined beam codebook for the phase shifter array. At this stage, no \ac{CSI} is required. A beam will be shaped by the phase shifter array and will be steered to determine the best directional beam depending on the target user received signal power. To simplify the beam steering, a codebook consisting of seven beam patterns is defined in Table \ref{tab:table_beam_codebook_hybrid_precoding}. The calculation of the relative phase offset \cite{phase_array_2005} between adjacent phase shifters is defined as
\begin{equation}
\varphi=\frac{360^{\degree}{\cdotp}d{\cdotp}sin\sigma}{\lambda},\label{eq:phase_shifter_relative_phase_offset}\end{equation}
where $\sigma$ is the steered beam angle in degrees, $\varphi$ is the relative phase offset between two adjacent phase shifters, $d$ is the distance between two adjacent phase shifters and $\lambda$ is the signal wavelength.

\begin{figure}[t!]
\begin{center}
\includegraphics[scale=0.25]{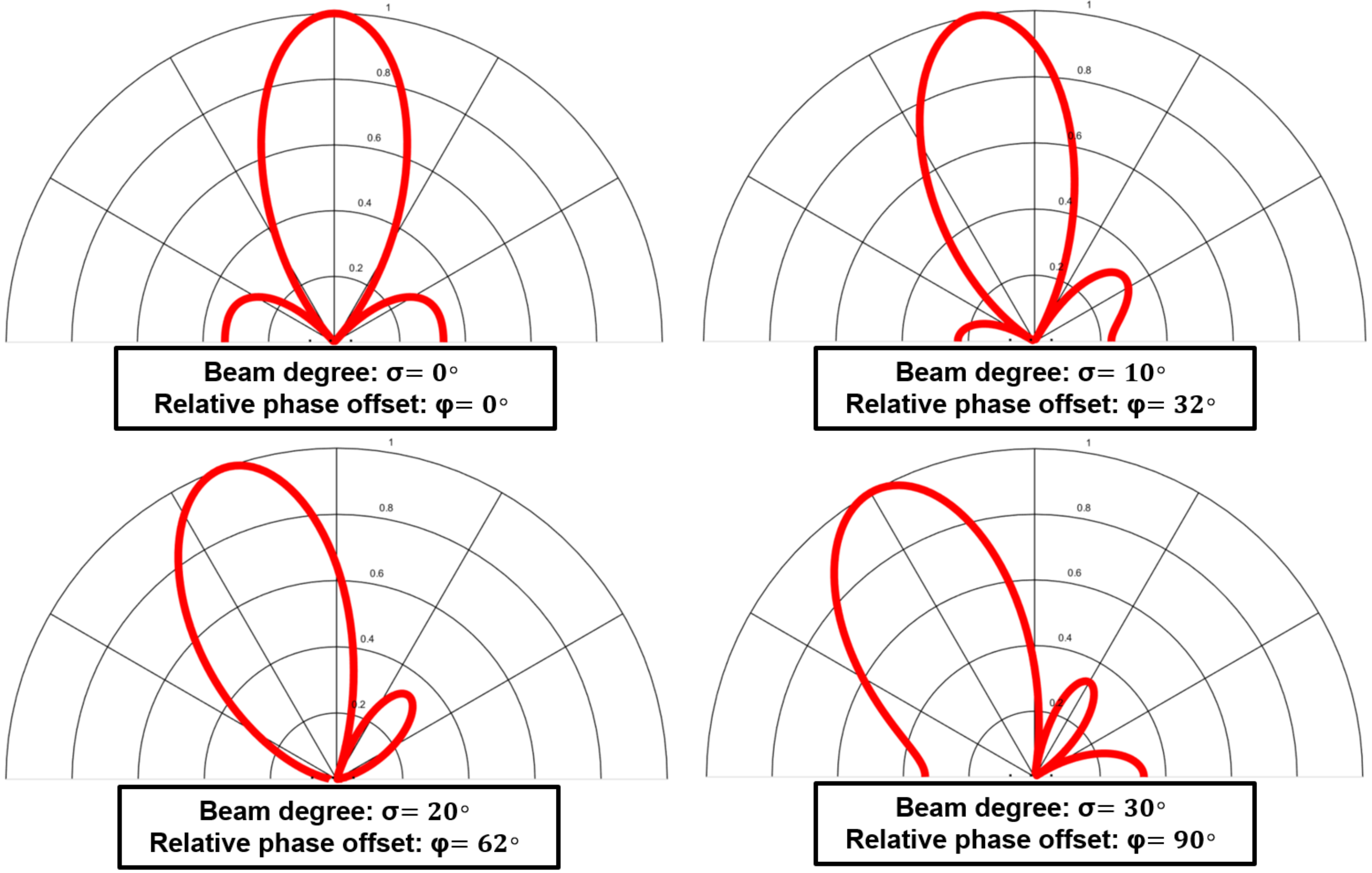}
\end{center}
\caption{Simulated beam patterns with predefined beam direction degrees. Three left-side hemisphere beam patterns and one zero-phase vertical beam pattern are illustrated and the right-side hemisphere beam patterns are the mirror symmetric copies of the left-side ones.}
\label{Fig:hybrid_analogue_beam_pattern}
\end{figure}

Since right-side hemisphere beam patterns are mirror symmetric copies of the left-side ones, then Fig. \ref{Fig:hybrid_analogue_beam_pattern} only illustrates three left-side hemisphere beam patterns and one zero-phase vertical beam pattern. All the beam patterns in Fig. \ref{Fig:hybrid_analogue_beam_pattern} are generated via three aligned omni-directional antennas, which are placed at $d=\lambda/2$ spacing.

\begin{table}[t!]
\caption{Beam pattern codebook for a three-antenna array per RF chain at $d=\lambda/2$ antenna spacing.}
\centering
\begin{tabular}{ccc}
\hline \hline
Pattern Index & Beam direction & Relative phase   \\[0.5pt] 
  &   degree $\sigma$  &  offset $\varphi$  \\[0.5pt] \hline 
0    & $0^{\degree}$   & $0^{\degree}$\\
1    & $10^{\degree}$  & $32^{\degree}$\\
2    & $20^{\degree}$  & $62^{\degree}$\\
3    & $30^{\degree}$  & $90^{\degree}$\\
4    & $-10^{\degree}$  & $32^{\degree}$\\
5    & $-20^{\degree}$  & $62^{\degree}$\\
6    & $-30^{\degree}$  & $90^{\degree}$\\ \hline \hline
\label{tab:table_beam_codebook_hybrid_precoding}
\end{tabular}
\end{table}

The optimal analogue beamforming protocol in \cite{hybrid_precoding_JSAC2009} divides the analogue beamforming into three steps to refine gradually, sharpen and track beams using a multi-resolution codebook. To simplify the analogue beamforming processing, a one-step beam search is applied in this work. The system would steer beams following the predefined patterns in Table \ref{tab:table_beam_codebook_hybrid_precoding} and select the best beam pattern with the highest target user received signal power. The beam refinement will be managed by the second stage MIMO digital precoding.

\begin{figure}[t!]
\begin{center}
\includegraphics[scale=0.45]{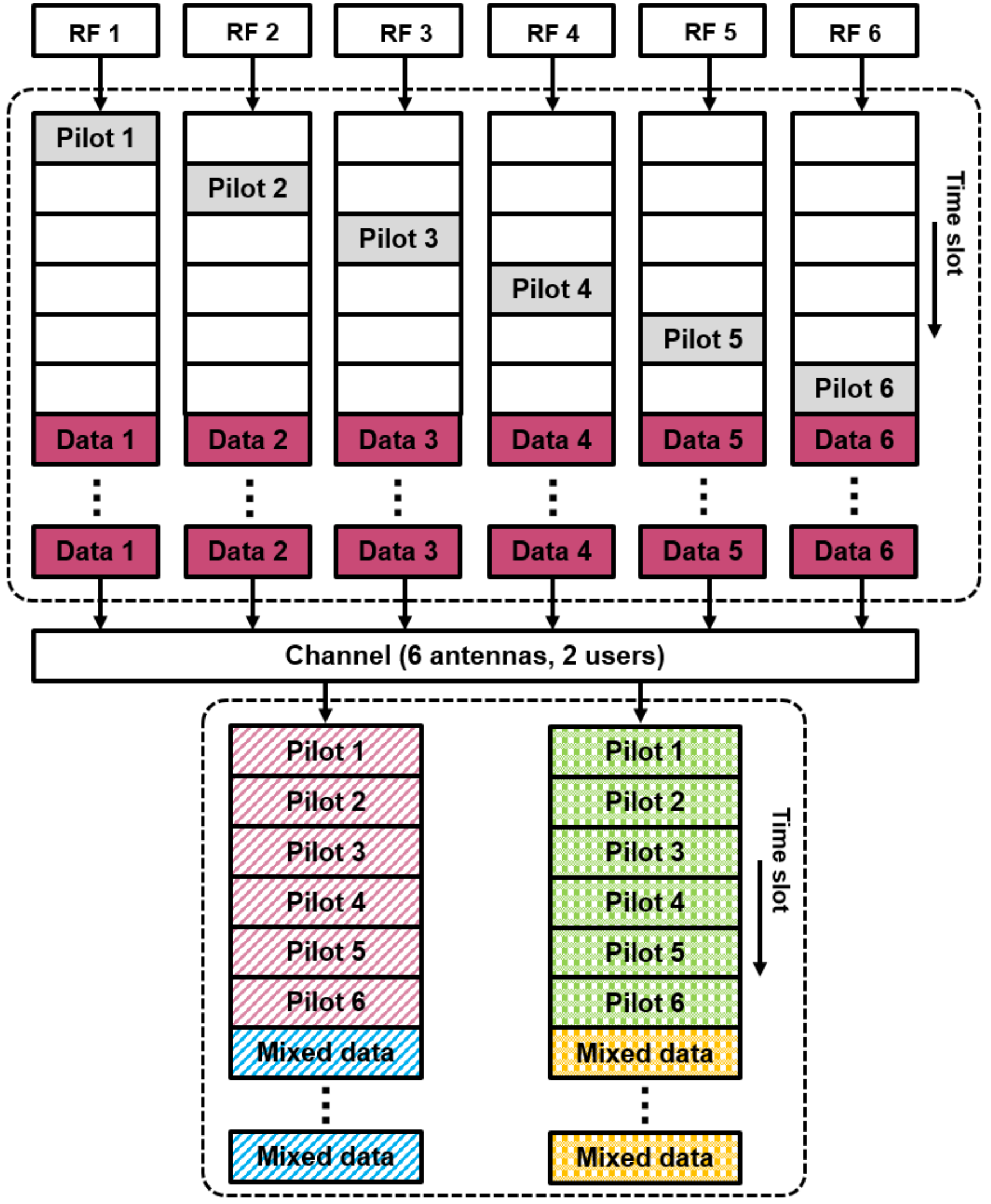}
\end{center}
\caption{Typical fully digital precoding staggered pilot frame structure. }
\label{Fig:full_digital_precoding_staggered_pilot_frame_structure}
\end{figure}

\begin{figure}[t!]
\begin{center}
\includegraphics[scale=0.45]{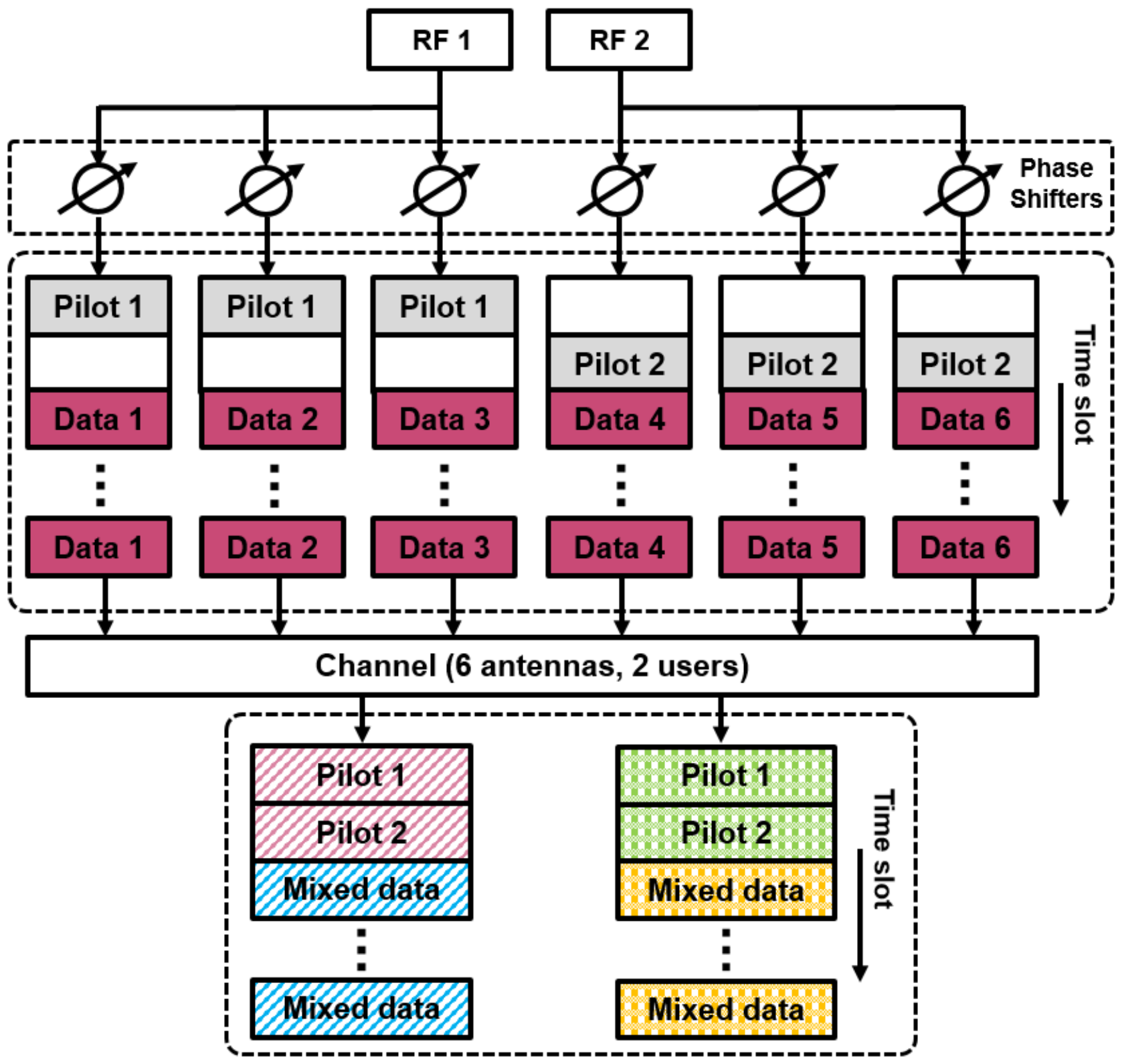}
\end{center}
\caption{Hybrid precoding staggered pilot frame structure. }
\label{Fig:hybrid_precoding_staggered_pilot_frame_structure}
\end{figure}

The configurations of analogue-domain signal phase and digital-domain signal amplitude and phase require iterative optimization \cite{hybrid_precoding_CM2014}, which is complex to practical systems. In addition, a practical, accurate and efficient channel estimation strategy is still an open research topic. Therefore, we aim to find the best analogue signal beam using a sub-optimal but a simple power detection approach according to the receiver side user feedback information. Prior to data transmission, pilot symbols are sent first to measure channel conditions. Pilot symbols from each RF chain are staggerly packed to avoid spatial signal overlapping interference.

\begin{figure*}[ht]
\begin{center}
\includegraphics[scale=0.3]{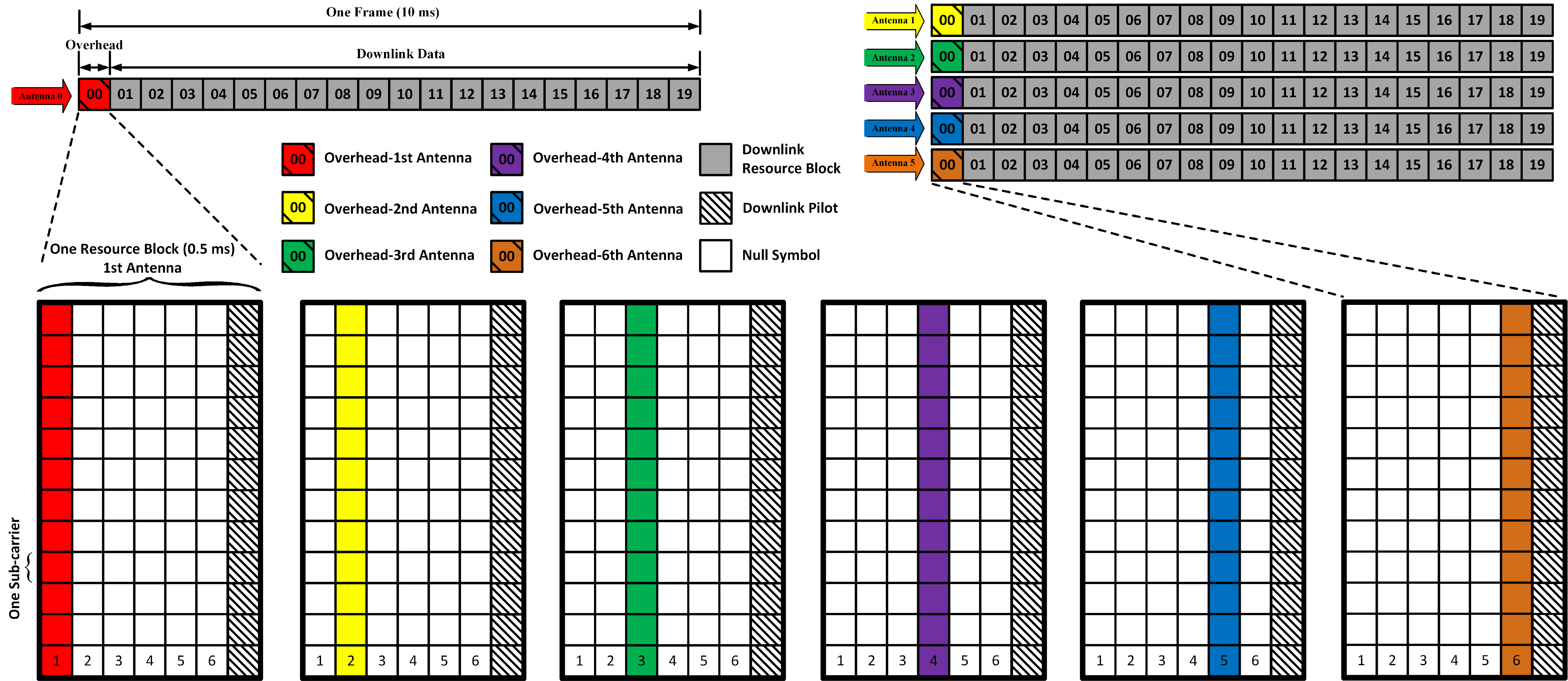}
\end{center}
\caption{Frame and resource block structure for the fully digital precoding system, reused from \cite{Tongyang_MU_MIMO_NB_IoT_2018}. One frame consists of 20 time slots and the first time slot is the overhead for downlink antenna spatial CSI estimation. The locations of overhead from different antennas are staggered to avoid inter-user interference. This example is for the six-RF-chain fully digital precoding system. For the two-RF-chain hybrid precoding system, each frame would have a shorter overhead indicating higher timing efficiency. }
\label{Fig:frame_RB_structure}
\end{figure*}

The pilot packing structures for the typical fully digital precoding system and the hybrid precoding system are illustrated in Fig. \ref{Fig:full_digital_precoding_staggered_pilot_frame_structure} and Fig. \ref{Fig:hybrid_precoding_staggered_pilot_frame_structure}, respectively. Assume the same number of transmitter antennas, Fig. \ref{Fig:full_digital_precoding_staggered_pilot_frame_structure} has to interleave six different pilot symbols in the time-domain, which is low efficient in timing resource utilization. Fig. \ref{Fig:hybrid_precoding_staggered_pilot_frame_structure} can efficiently improve the timing resource utilization by means of sending only two pilot symbols due to the signal duplications of power splitters and signal beamforming of phase shifters. In this case, only two pilot symbols are sufficient even though six antennas are employed. It should be noted that the staggered structure is only for pilot symbols since the inter-user interference-free pilot transmission strategy enables accurate channel estimations and further helps the beamforming selection. For non-pilot data transmission, different RF chains would send data at the same time even with inter-user interference \cite{Tongyang_MU_MIMO_NB_IoT_2018}. With the first-step channel estimation, the spatial interference caused by non-pilot data from different antennas would be removed. Therefore, in terms of pilot timing resources, the hybrid precoding architecture has a higher timing efficiency than that of the fully digital precoding counterpart. However, both architectures have the same timing efficiency when non-pilot data is transmitted.

The time-frequency frame of the pilot and non-pilot symbols allocation is reused from \cite{Tongyang_MU_MIMO_NB_IoT_2018} and illustrated in Fig. \ref{Fig:frame_RB_structure} where the six-RF-chain fully digital precoding is realized, showing that all six antennas transmit pilot symbols at the same set of frequencies but each in a different symbol time location. Thus, time-domain inter-user interference can be avoided. For non-pilot symbol transmission, the six antennas will transmit data at the same time and on the same set of frequencies. For the two-RF-chain hybrid precoding architecture, two antenna frames would be used instead of the six frames in Fig. \ref{Fig:frame_RB_structure}. In this case, two staggered pilot symbols would be sufficient as the overhead for the MIMO spatial interference estimation. Therefore, timing efficiency is improved.

A single beam would be generated after using three phase shifters on each RF chain. Therefore, we define in \eqref{eq:X_matrix} equivalent pilot data $p_1$ for the first beam and $p_2$ for the second beam.
\begin{equation}
\mathbf{P} =
 \begin{bmatrix}
  p_{1} & 0  \\
  0     & p_{2}
 \end{bmatrix}.
\label{eq:X_matrix}\end{equation}

To achieve an optimal beam, the phase shifters have to be accurately tuned according to the pre-defined beam pattern codebook in Table \ref{tab:table_beam_codebook_hybrid_precoding}. However, due to the imperfect beam characteristics such as the beam width and side lobes, inter-user interference is not completely avoided. This results in a full channel matrix, instead of a diagonal matrix, even when hybrid precoding is used.

One benefit of using the hybrid precoding architecture is that the effective channel $\mathbf{H}$ is simplified from a $2\times{6}$ matrix (in Fig. \ref{Fig:full_digital_precoding_staggered_pilot_frame_structure}) to a $2\times{2}$ matrix (in Fig. \ref{Fig:hybrid_precoding_staggered_pilot_frame_structure}). 
\begin{equation}
\mathbf{H} =
 \begin{bmatrix}
  h_{11} & h_{12} \\
  h_{21} & h_{22}
 \end{bmatrix}.
\label{eq:H_matrix}\end{equation}

Thus, the received pilot signals at each user, through MIMO channels and \ac{AWGN} $z$, are given by

\begin{equation}
 \begin{bmatrix}
  y_{11} & y_{12} \\
  y_{21} & y_{22}
 \end{bmatrix}=
 \begin{bmatrix}
  h_{11} & h_{12} \\
  h_{21} & h_{22}
 \end{bmatrix}\times
 \begin{bmatrix}
  p_{1} & 0  \\
  0     & p_{2}
 \end{bmatrix}+
 \begin{bmatrix}
  z_{1}  \\
  z_{2}
 \end{bmatrix},
\label{eq:Y_matrix}\end{equation} 
where $y_{11}$ and $y_{12}$ are the received symbols at the first user in the first time period and the second time period, respectively. The two symbols $y_{21}$ and $y_{22}$ are received at the second user. The criterion of selecting the best beam pattern depends on the power measurement at the receiver. The staggered pilot structure not only measures the target user signal power, but also the interference signal power coming from the other user. Since pilots from each RF chain are time orthogonal, each user would be able to measure received signal power without inter-user interference. Therefore, the first user only takes into account $P_{11}$ and the second user measures $P_{22}$. 
\begin{eqnarray}P_{11}=|y_{11}|^{2},\label{eq:P11_power}\end{eqnarray}
\begin{eqnarray}P_{22}=|y_{22}|^{2}.\label{eq:P22_power}\end{eqnarray}

The pattern index associated with the best beam is selected according to the highest user signal power as the following.
\begin{eqnarray}PI_1=\arg\max_{{PI_1}\in M}P_{11},\label{eq:PI_1_measure}\end{eqnarray}
\begin{eqnarray}PI_2=\arg\max_{{PI_2}\in M}P_{22},\label{eq:PI_2_measure}\end{eqnarray}
where $M$ is the beam patterns in Table \ref{tab:table_beam_codebook_hybrid_precoding}, $PI_1$ indicates the beam pattern index for the first user and $PI_2$ indicates the beam pattern index for the second user. The pattern index, leading to the highest target user power, is the final solution.

\subsection{Stage-II: MIMO Digital Precoding}

The first stage analogue precoding converts a high-dimension RF domain system to a low-dimension baseband system. This stage is to fine-tune the beam in digital-domain by modifying both signal amplitude and phase. Based on the amplitude and phase of received pilots, robust channel estimation is realized for the digital precoding with the channel coefficients estimation $\mathbf{\hat{H}}$ in the following.
 
\begin{equation}
\mathbf{\hat{H}} =
 \begin{bmatrix}
  \hat{h}_{11} & \hat{h}_{12} \\
  \hat{h}_{21} & \hat{h}_{22}
 \end{bmatrix}=
 \begin{bmatrix}
  y_{11}/p_{1} & y_{12}/p_{2} \\
  y_{21}/p_{1} & y_{22}/p_{2}
 \end{bmatrix}.
\label{eq:H_estimate_matrix}\end{equation} 

The problem of inter-user pilot interference is avoided by employing the newly developed staggered pilot transmission, shown in Fig. \ref{Fig:full_digital_precoding_staggered_pilot_frame_structure} and Fig. \ref{Fig:hybrid_precoding_staggered_pilot_frame_structure}. This allows accurate and real time channel estimations using \eqref{eq:H_estimate_matrix} and alleviates the need of using the impractical complex \ac{ML} estimation. In addition, the first stage analogue precoding already shapes the beam in the space-domain, resulting in better channel conditions with less interference in the digital precoding stage. Both factors indicate that the simple estimation method in \eqref{eq:H_estimate_matrix} would be sufficient.

It should be noted that unlike the first stage analogue precoding, the digital precoding can modify both the signal amplitude and phase. The operation relies on \ac{ZF} and the effective low-dimension digital-domain \ac{CSI} $\mathbf{\hat{H}}$. The MIMO digital precoding matrix $\mathbf{D}$ is computed as

\begin{equation}
\mathbf{D}=\mathbf{\hat{H}}^H(\mathbf{\hat{H}}\mathbf{\hat{H}}^H)^{-1}=
 \begin{bmatrix}
  d_{11} & d_{12} \\
  d_{21} & d_{22}
 \end{bmatrix}. \label{eq:digital_precoding}\end{equation}

Therefore, the non-pilot data signals, prior to the MIMO antenna transmission, is precoded as 

\begin{equation}
 \begin{bmatrix}
  \bar{X}_1  \\
  \bar{X}_2
 \end{bmatrix}=
 \begin{bmatrix}
  d_{11} & d_{12} \\
  d_{21} & d_{22}
 \end{bmatrix}\times
 \begin{bmatrix}
  \tilde{X}_1  \\
  \tilde{X}_2
 \end{bmatrix},
\label{eq:precoding_tx_MIMO}\end{equation} 
where $\tilde{X}_1$ and $\tilde{X}_2$ are the non-pilot original data streams generated at the first RF chain and the second RF chain, respectively. Each of them includes multiple OFDM/SEFDM symbols as is shown in Fig. \ref{Fig:frame_RB_structure}. The precoded data streams $\bar{X}_1$ and $\bar{X}_2$ will be transmitted at the same time and frequency.

Therefore, the staggered pilot symbols can simultaneously assist to find the best analogue beamforming pattern for the Stage-I analogue precoding and further obtain inter-user interference free channel estimation coefficients for the Stage-II MIMO digital precoding.

\subsection{Stage-III: Waveform Precoding}

Waveform precoding is to pre-equalize the interfered SEFDM signals using the pre-defined correlation matrix $\mathbf{C}$, which consists of elements $c_{m,n}$ derived from \eqref{eq:element_corr}. The correlation matrix is deterministic once the signal waveform is generated. Therefore, the waveform precoding is straightforward by multiplying raw signals with the inverse of the correlation matrix $\mathbf{C}$ and no \ac{CSI} is needed at this stage. The waveform precoding matrix $\mathbf{W_p}$, based on \ac{ZF}, is thus defined as
\begin{equation}
\mathbf{W_p}=\mathbf{C}^H(\mathbf{C}\mathbf{C}^H)^{-1}. \label{eq:waveform_precoding}\end{equation}

The waveform precoding, conducted on $S$ in \eqref{eq:Tx_signal_model}, gives a new expression in the following.
\begin{equation}
\bar{S}=\mathbf{W_p}S.\label{eq:precoding_tx_waveform}\end{equation}

The \ac{ZF} based precoding would inevitably amplify any deviations coming from the first two precoding stages. This indicates the importance of accurate analogue and digital precoding. It should be noted that the precoding matrix $\mathbf{W_p}$ can be optimized using other algorithms such as \ac{CI} \cite{Tongyang_paris_INFOCOM_2019} rather than ZF, which would be possible solutions of mitigating the deviation sensitivity issue. 

\section{Multiuser MIMO-SEFDM System Prototyping}

The experiment is conducted in an indoor laboratory, which is 4 meters wide and 9 meters long, at a carrier frequency $f_{RF}$=2.4 GHz. The experiment has a static indoor channel environment due to fixed user and base station locations. Therefore, slow beam sweeping is adequate. This setup covers a number of realistic scenarios where IoT devices largely remain in fixed locations after initial deployment. In addition, due to the nature of narrow bandwidth of IoT signals, frequency selective channel impairments are negligible \cite{Erik_book_4G}.

This work focuses on small size systems of six base station antennas and two receiver users taking into account realistic scenarios and limitations. Firstly, IoT communications prefer low-power low-complexity system architectures. In addition, there are no requirements for multiple-antenna system architectures for IoT applications. This experiment work therefore represents a significant step forward to test up to six antennas at the transmitter for functionality verifications. Secondly, this work aims to explore the physical layer signal transmissions and therefore skip any additional layer protocols such as time- frequency- code- domain multiple access and link scheduling. Thus, the number of accessed users in this experiment is functionally limited. Accordingly, the findings of our study can be applied simply to multiple such domains, by employing additional layers of scheduling and time-frequency division. This can tremendously extend the potentials of the proposed signal format, to achieve the massive connectivity. Thirdly, to have a better backward compatibility with our previous experiment testbeds \cite{Tongyang_MU_MIMO_NB_IoT_2018, Tongyang_paris_INFOCOM_2019}, the same system scale is configured in this work for fair comparisons.

\begin{figure*}[t!]
\begin{center}
\includegraphics[scale=0.48]{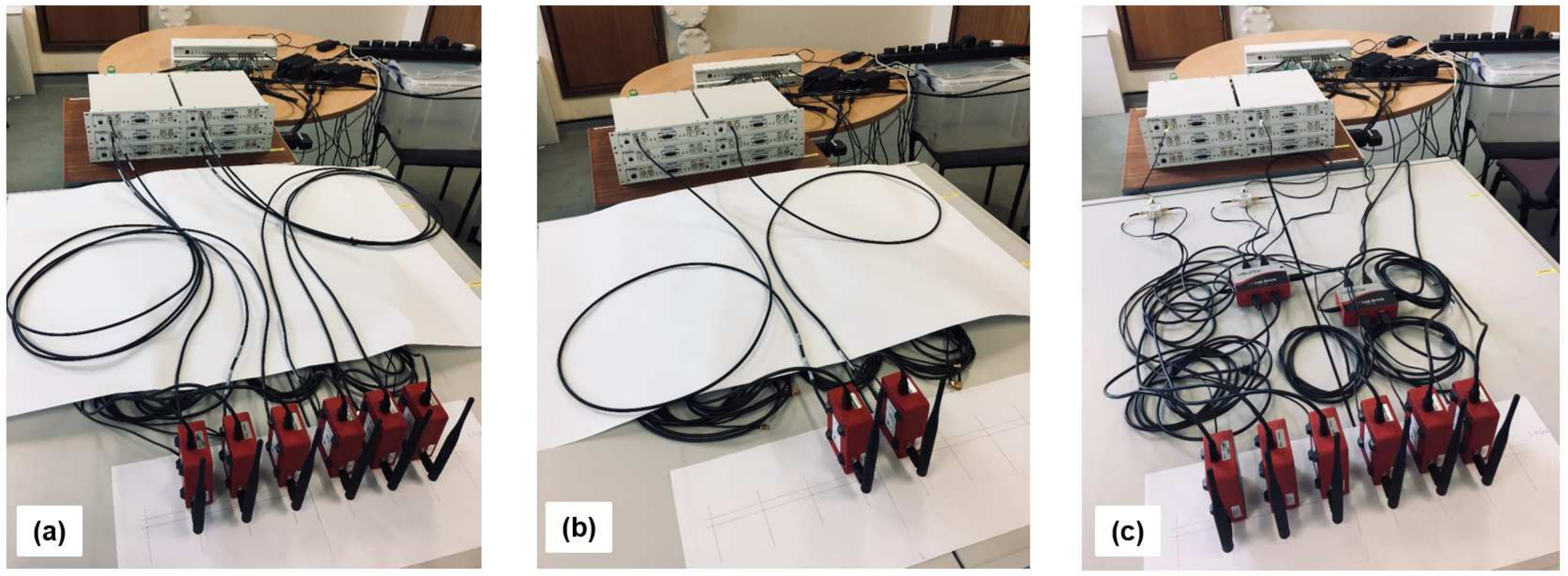}
\end{center}
\caption{Testbed setups of different precoding architectures. (a) Fully digital precoding (FDP-I) system with $N_{RF}$=6 RF chains and $N_{TX}$=6 transmitter antennas. Phase shifters here are only for antenna alignment with no phase controls. (b) Fully digital precoding (FDP-II) system with $N_{RF}$=2 RF chains and $N_{TX}$=2 transmitter antennas. Phase shifters here are only for antenna alignment with no phase controls. (c) Three-stage hybrid analogue-digital precoding (HP) system with $N_{RF}$=2 RF chains and $N_{TX}$=6 transmitter antennas. Phase shifters here are both for antenna alignment and phase controls.}
\label{Fig:Hybrid_beamforming_testbed}
\end{figure*}

The aim of this work is to show the interference mitigation capability of using hybrid analogue-digital precoding in MIMO-SEFDM signals. To have a comprehensive comparison, two additional fully digital precoding systems are designed and implemented with system architecture configurations in Table \ref{tab:table_system_architecture_hybrid_precoding}. The practical testbed setups of the three systems are demonstrated in Fig. \ref{Fig:Hybrid_beamforming_testbed}.

\begin{table}[t!]
\caption{Experiment testbed architecture configurations}
\centering
\begin{tabular}{llll}
\hline \hline
$\mathbf{Parameter}$ & $\mathbf{FDP-I}$ & $\mathbf{FDP-II}$ & $\mathbf{HP}$ \\[0.5pt] \hline 
No. of RF chains    & 6 & 2 & 2\\
No. of transmitter antennas    & 6 & 2 & 6\\
No. of power splitters    & 0 & 0 & 2\\
No. of phase shifters    & 0 & 0 & 6\\ \hline \hline
\label{tab:table_system_architecture_hybrid_precoding}
\end{tabular}
\end{table}

The experiment employs Vaunix LPS-402 programmable phase shifter \cite{Vaunix_phase_shifter_ref} for two reasons. First, it has a high flexibility to align antennas at specific spacing. Second, its programmable feature and its high phase increment resolution (one degree) are vital to the design of hybrid analogue-digital precoding systems. The required experimentation equipment and devices are listed in Table \ref{tab:hybrid_precoding_required_devices} and the key signal parameters are summarized in Table \ref{tab:table_waveform_hybrid_precoding}.

\begin{table}[t!]
\caption{Required devices for experiments}
\begin{center}
\begin{tabular}{llll}
\hline \hline
Item  & Description & Quantity \\ \hline 
1   & NI USRP-2953R  &  6 \\ 
2   & CPS-8910 cabled PCIE switch box x4, 10-Port  & 1  \\ 
3   & MXI-Express cable, Gen 1 x4, Copper & 6  \\ 
4   & MXI-Express cable, Gen 2 x8, Copper & 1 \\ 
5   & CDA-2990 8-channel clock distribution OctoClock  &  1  \\ 
6   & Transmitter omni-directional antenna  &  6  \\ 
7   & User omni-directional antenna  &  2  \\ 
8   & Control computer  &  1  \\ 
9   & Vaunix LPS-402 phase shifter  &  6  \\ 
10  & Power splitter  &  2  \\ \hline \hline
\end{tabular}
\label{tab:hybrid_precoding_required_devices}
\end{center}
\end{table}

\begin{table}[t!]
\caption{Experiment signal specifications}
\centering
\begin{tabular}{lll}
\hline \hline
$\mathbf{Parameter}$ & $\mathbf{OFDM}$ & $\mathbf{SEFDM}$ \\[0.5pt] \hline 
RF center frequency (GHz) & 2.4 & 2.4 \\ 
Sampling frequency (MHz) & 1.92 & 1.92 \\ 
IFFT sample length & 128 & 128 \\ 
No. of guard band sub-carriers & 58 & 58 \\ 
No. of data sub-carriers & 12 & 12 \\ 
No. of cyclic prefix samples & 10 & 10 \\ 
Sub-carrier bandwidth (kHz) & 15 & 15 \\ 
Bandwidth compression factor $\alpha$ & 1 & 0.9\\ 
Sub-carrier spacing (kHz) & 15 & 15$\times{\alpha}$ \\ 
Bandwidth (kHz) & 180 & 180${\times}\alpha$ \\ 
Beam degree range  & $-30^{\degree}$:\ $30^{\degree}$ & $-30^{\degree}$:\ $30^{\degree}$ \\
Modulation formats & {4QAM, 16QAM}  & {4QAM, 16QAM} \\
Peak spectral efficiency (bit/s/Hz) & {2, 4}  & {2/$\alpha$, 4/$\alpha$} \\  \hline \hline
\label{tab:table_waveform_hybrid_precoding}
\end{tabular}
\end{table}

\begin{figure}[t!]
\begin{center}
\includegraphics[scale=0.135]{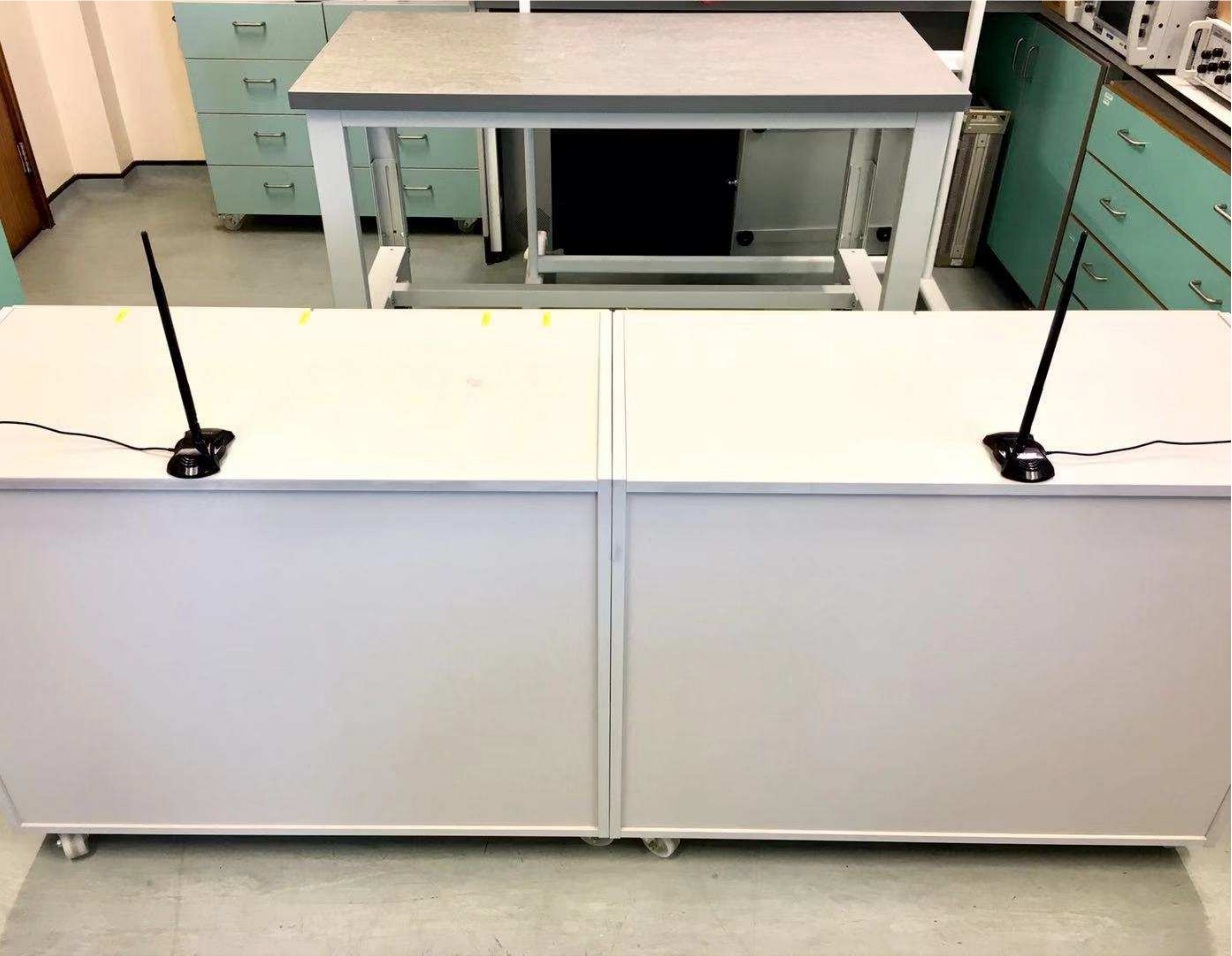}
\end{center}
\caption{Receiver side users. $N_{user}$=2 and each user has $N_{RX}$=1 antenna. }
\label{Fig:hybrid_precoding_users}
\end{figure}

Two users are placed in front of the transmitter antenna array with a \ac{LOS} path and each user is equipped with $N_{RX}$=1 antenna. The distance between the users and the antenna array is flexible and the user location is flexible as well. In this experiment as demonstrated in Fig. \ref{Fig:hybrid_precoding_users}, we separate two users (antennas) by 1.1 m and their distance to the base station is set to 2 m. A beam shaped by phase shifters, according to the beam codebook in Table \ref{tab:table_beam_codebook_hybrid_precoding}, is steered and the best beam pattern is selected based on the highest target user received signal power.

\subsection{Fully Digital Precoding Testbed: FDP-I}

In the fully digital precoding experiment as demonstrated in Fig. \ref{Fig:Hybrid_beamforming_testbed}(a), the phase shifters are only used for antenna alignment. Therefore, to remove phase interference, the initial phases of the phase shifters are all set to zero. An array of six zero-degree phase shifters are directly connected to six RF chains via SMA cables. The RF chain is provided by a commercialized software defined radio (SDR) platform USRP-RIO 2953R \cite{NI_SDR_ref}. An omni-directional antenna is connected to each phase shifter with the spacing of $\lambda/2$. Six USRP-RIO 2953R devices are used to provide six independent RF chains. A CDA-2990 8-channel clock distribution OctoClock module is connected to synchronize six USRPs by providing 10 MHz reference signals and pulse per second (PPS) signals. The baseband digital signal is generated in software LabVIEW and distributed to six USRPs via a cabled PCI-express switch box CPS-8910. Digital signals are converted to analogue signals in each USRP and up converted to 2.4 GHz carrier frequency.

\subsection{Fully Digital Precoding Testbed: FDP-II}

Since RF chains are expensive to be implemented in practical systems, we design and test a two-RF-chain system in Fig. \ref{Fig:Hybrid_beamforming_testbed}(b). The system architecture is similar to the fully digital precoding system in Fig. \ref{Fig:Hybrid_beamforming_testbed}(a) except for the number of RF chains and transmitter antennas. The antenna spacing is still half wavelength but only two USRPs are connected in this case. The initial phase is set to zero in each phase shifter to avoid any phase interference.

\subsection{Hybrid Analogue-Digital Precoding Testbed: HP}

The hybrid precoding system following a sub-connected architecture is demonstrated in Fig. \ref{Fig:Hybrid_beamforming_testbed}(c). For each RF chain, three phase shifters are placed in alignment with half wavelength spacing. Power splitters are connected between phase shifters and RF chains. In this case, each RF chain feeds the same signal to three phase shifters via power splitters. The initial phase of each phase shifter can be tuned to realize analogue precoding functions.

The system sweeps the beam by changing the phase in each phase shifter. Therefore, the user side received signal power would be variable. The beam, which enables the highest receiver side signal power, is the best one. After one sweeping cycle, the phase shifters are configured to the optimal phase parameters. It should be noted that an ideal narrow beam is not easily generated in practice due to the limited number of phase shifters and the low radio carrier frequency. Since the experiment works at 2.4 GHz radio frequency and only three antennas are connected to each RF chain, the inter-user interference is inevitable. In addition, the shaped beam from the phase shifter array has sidelobes, which causes interference to neighbouring users as well. Therefore, the second stage digital precoding is necessary to jointly fine-tune to enhance user diversity gains and multiplexing quality.

\section{Measured Results}

We measure constellation patterns, \ac{EVM}, \ac{BER}, \ac{SE} and {\ac{EE}} as performance indicators in this experiment. Previous work \cite{Tongyang_wincom2017} evaluated 16QAM-SEFDM with small bandwidth compression factors. The results in \cite{Tongyang_wincom2017} revealed that by properly tuning bandwidth compression factors, 16QAM-SEFDM can outperform 32QAM-OFDM and 64QAM-OFDM of the same spectral efficiency. However, this is at the cost of complex receiver side signal processing, which is not practical in our proposed low-complexity IoT scenarios. In this work, 4QAM and 16QAM modulation formats are used in the experiments to evaluate both OFDM and SEFDM signals transmission with transmitter side precoding processing.

\begin{figure}[t!]
\begin{center}
\includegraphics[scale=0.43]{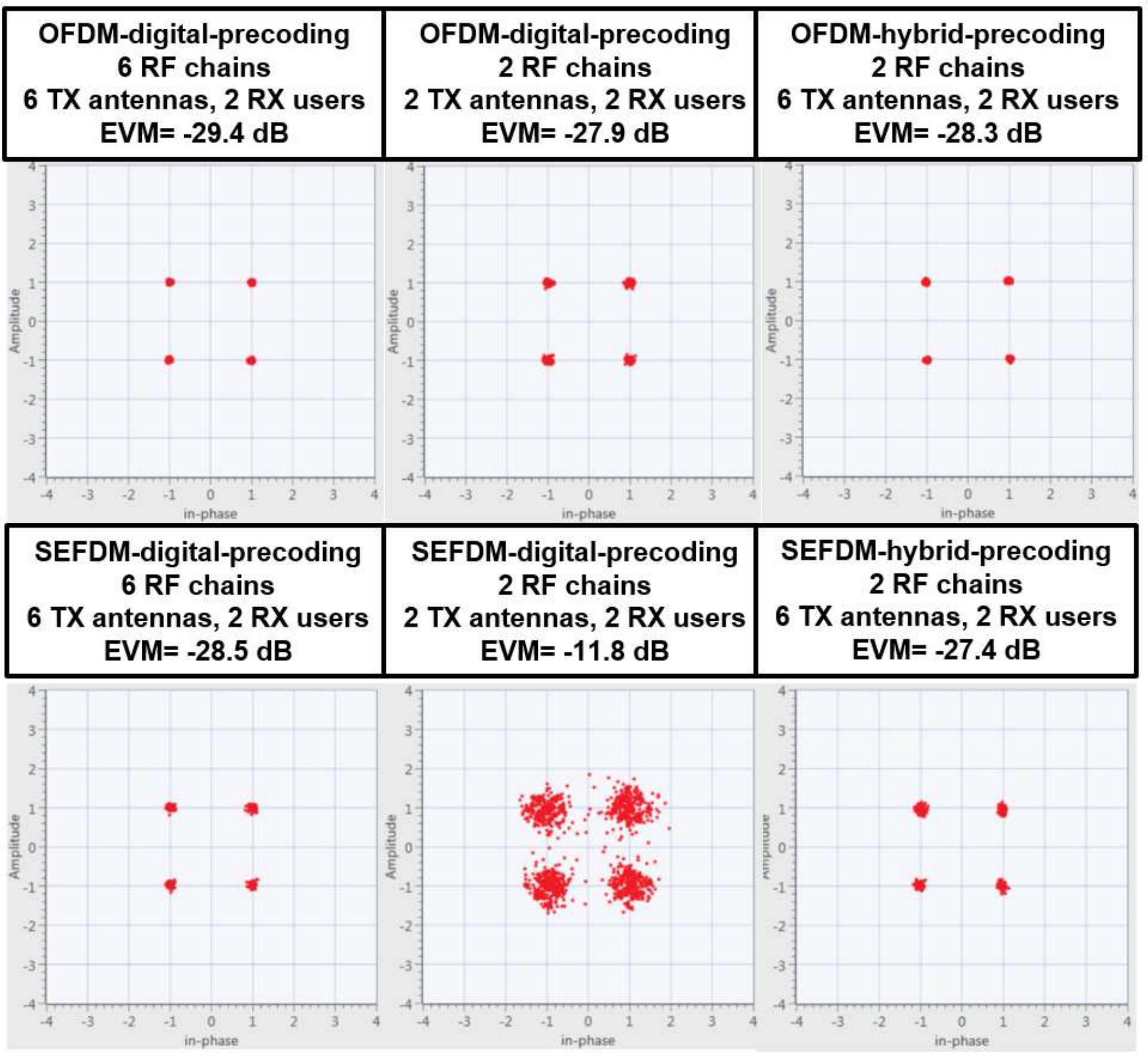}
\end{center}
\caption{Measured 4QAM constellation and EVM. }
\label{Fig:hybrid_precoding_results_4QAM_constellation}
\end{figure}

As is shown in Fig. \ref{Fig:hybrid_precoding_results_4QAM_constellation}, the number of RF chains and transmitter antennas have no apparent \ac{EVM} difference for 4QAM modulated OFDM signals. There could be multiple reasons leading to this result such as the small number of users, good channel conditions and the low level modulation format. It could be concluded that two antennas are sufficient for pure digital precoding when two users are considered in the orthogonal signal waveform \ac{OFDM}. However, the number of RF chains and transmitter antennas have great effects on the performance for 4QAM modulated SEFDM signals. The reason may come from the third stage waveform precoding since a precoding matrix is multiplied to equalize the SEFDM waveform \ac{ICI}. Therefore, any deviations from the first two precoding stages would be amplified at the thrid stage. The SEFDM system of two RF chains is more likely to have accuracy deviations and would cause performance degradation amplifications. Extra flexibility of magnitude and phase adjustment is required for the ICI mitigation, which indicates more RF chains and transmitter antennas. With more antennas connected to each RF chain via phase shifters, better performance is achieved since the beam shaping from the first stage analogue precoding catalyze more accurate amplitude and phase tuning at the second stage digital precoding. In summary, based on our indoor experiment, for 4QAM modulated OFDM signals, hybrid precoding has no obvious performance gains when compared with pure digital precoding since sub-carriers are orthogonally packed. However, hybrid precoding has great performance improvement on 4QAM modulated SEFDM signals due to the deviation sensitivity of waveform precoding.

\begin{figure}[t!]
\begin{center}
\includegraphics[scale=0.42]{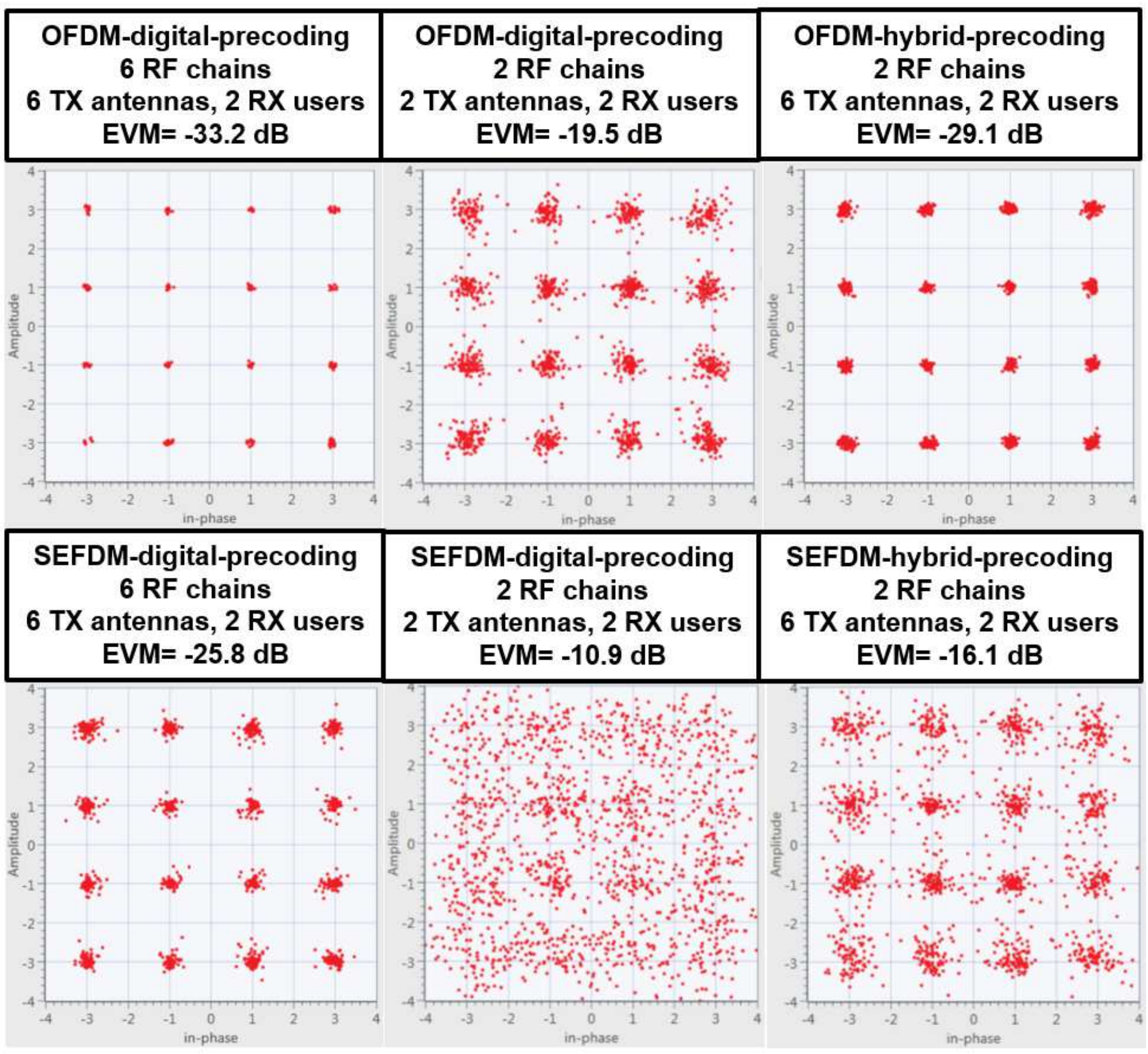}
\end{center}
\caption{Measured 16QAM constellation and EVM. }
\label{Fig:hybrid_precoding_results_16QAM_constellation}
\end{figure}

For higher modulation formats such as 16QAM, different results are observed in Fig. \ref{Fig:hybrid_precoding_results_16QAM_constellation}. The fully digital precoding OFDM system of six RF chains achieves a reasonable performance while reducing it to two RF chains resulting in evident performance loss. This may comes from the high constellation density of 16QAM and therefore the two-RF-chain system is not accurate sufficient to fine-tune signal amplitude and phase at the second stage digital precoding. With the help of extra antennas and phase shifters while maintaining two RF chains, the performance is apparently improved approaching the fully digital precoding system of six RF chains. In terms of SEFDM, the performance of six RF chains is slightly worse than that of OFDM due to the joint effect of dense constellation and waveform \ac{ICI}. Since the SEFDM system of six RF chains starts to show unreliable performance, cutting more RF chains would further degrade performance, which is shown as the scattered SEFDM constellation. With phase shifters and more antennas connected, the performance is improved but it is still worse than the OFDM.

In terms of performance improvement, hybrid precoding is more efficient for SEFDM since the ICI waveform precoding is very sensitive to the deviations caused by initial precoding stages and is also sensitive to multi-level modulation formats due to higher constellation density.

\begin{figure}[t!]
\begin{center}
\includegraphics[scale=0.49]{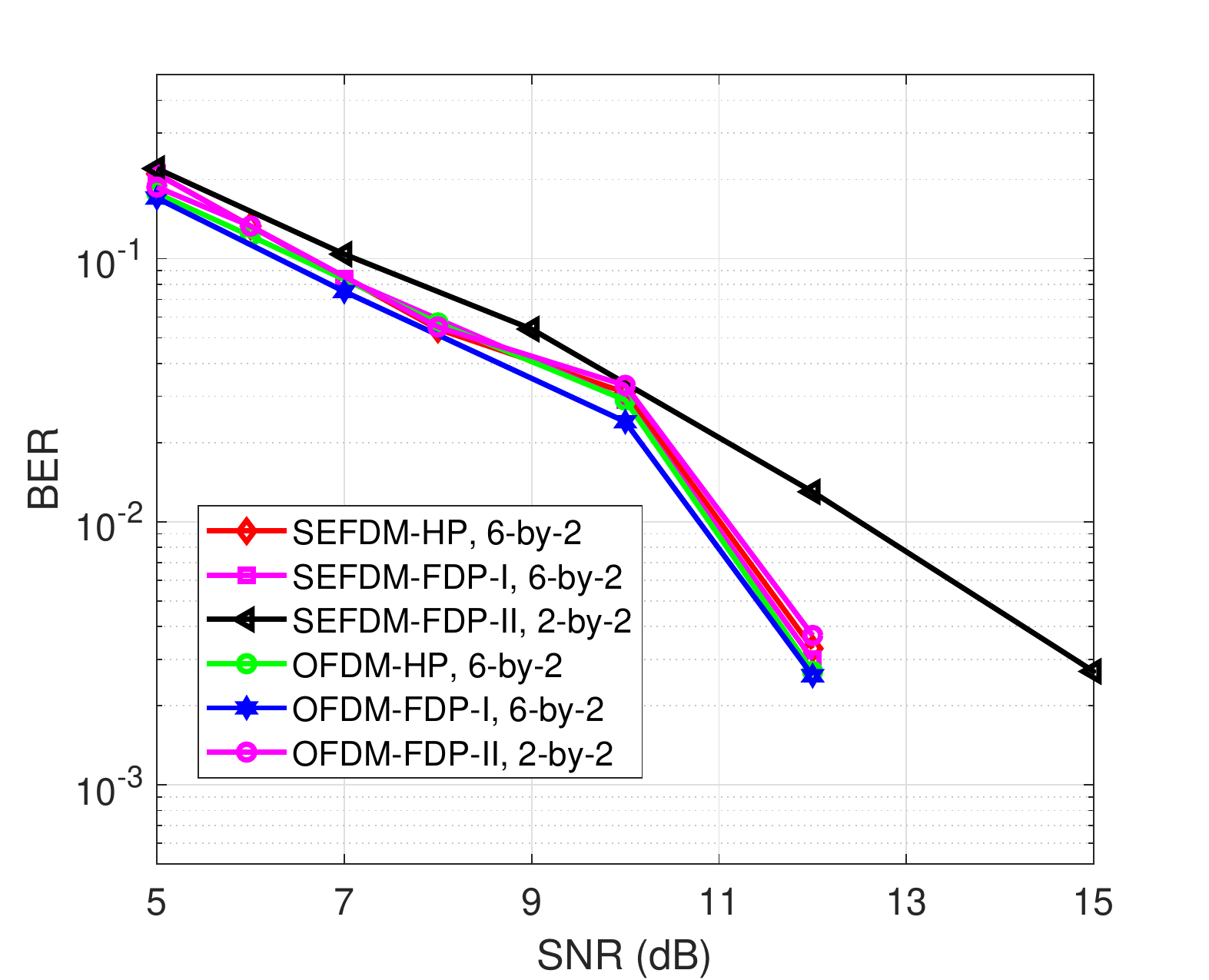}
\end{center}
\caption{Measured 4QAM BER. }
\label{Fig:Hybrid_precoding_SEFDM_BER_4QAM}
\end{figure}

\begin{figure}[t!]
\begin{center}
\includegraphics[scale=0.49]{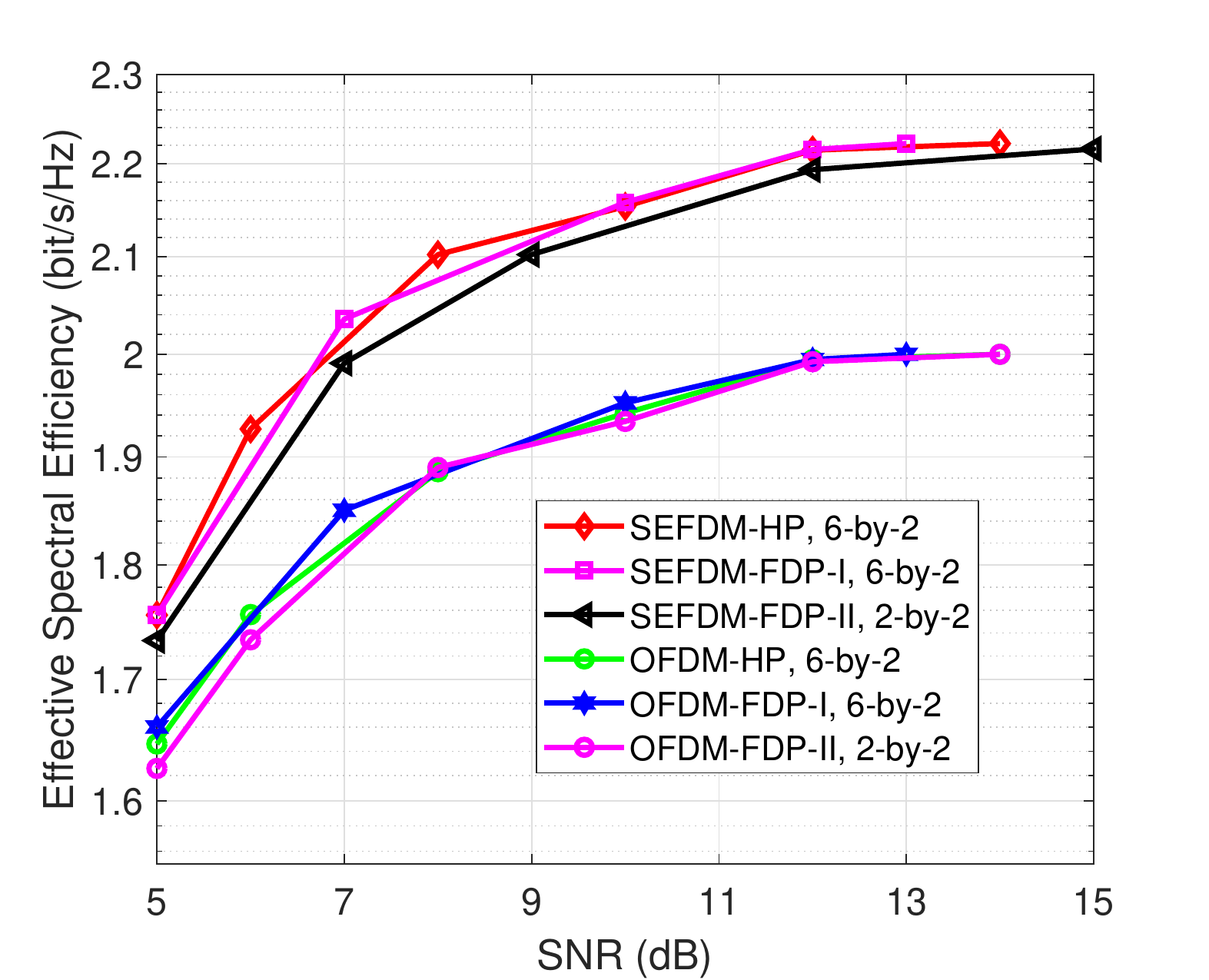}
\end{center}
\caption{Measured 4QAM spectral efficiency. }
\label{Fig:Hybrid_precoding_SEFDM_SE_4QAM}
\end{figure}

In addition to constellation diagrams, BER and effective spectral efficiency are also presented in this section. The BER performance, shown in Fig. \ref{Fig:Hybrid_precoding_SEFDM_BER_4QAM}, reveals that the fully digital precoding OFDM has the best performance while other precoding systems are slightly worse but are approaching the best performance. However, the fully digital precoding SEFDM system of two RF chains has a great performance loss compared with other systems. This is due to the aforementioned deviation amplification at the waveform precoding stage. It indicates that a higher number of RF chains or antennas have to be equipped.

Effective spectral efficiency is defined as the ratio between effective bit rate \cite{Tongyang_MU_MIMO_NB_IoT_2018} and signal spectral bandwidth. The effective spectral efficiency is expressed as
\begin{equation}
SE_e=R_e/B_e=(1-BER)\times{f_s}\times{log_2{O}}\times{(N_d/N)}/B_e,\label{eq:throughput_effective_calculation}\end{equation}
where $SE_e$ is the effective spectral efficiency, $R_e$ is the effective bit rate and $B_e$ is the effective spectral bandwidth; $B_e=B$ for OFDM signals and $B_e=\alpha\times{B}$ for SEFDM signals. $f_s$ is sampling rate, $O$ is constellation cardinality, $(1-BER)$ indicates the probability of non-error received bits, $N_{d}$ is the number of data sub-carriers and $N$ is the total number of sub-carriers. {The sampling rate $f_s$ indicates the number of samples can be delivered per second. In practical systems, null sub-carriers are added on both sides of data sub-carriers for the purpose of guard band protection. Therefore, the effective symbol rate is determined by the ratio between the number of data sub-carriers $N_{d}$ and the total number of sub-carriers $N$. Furthermore, to obtain the correctly transmitted symbol rate, $(1-BER)$ has to be considered as well. In addition, the conversion from symbol rate to bit rate is related to the modulation constellation cardinality $O$. Basically, higher data rates would be achieved with higher sampling rates, higher data sub-carrier percentages, higher modulation formats and lower BER.}

In terms of effective spectral efficiency in Fig. \ref{Fig:Hybrid_precoding_SEFDM_SE_4QAM}, SEFDM systems with all RF chain architectures show higher effective spectral efficiency levels than OFDM systems. The higher number of RF chains or antennas, the faster for SEFDM to reach the maximum spectral efficiency.

\begin{figure}[t!]
\begin{center}
\includegraphics[scale=0.49]{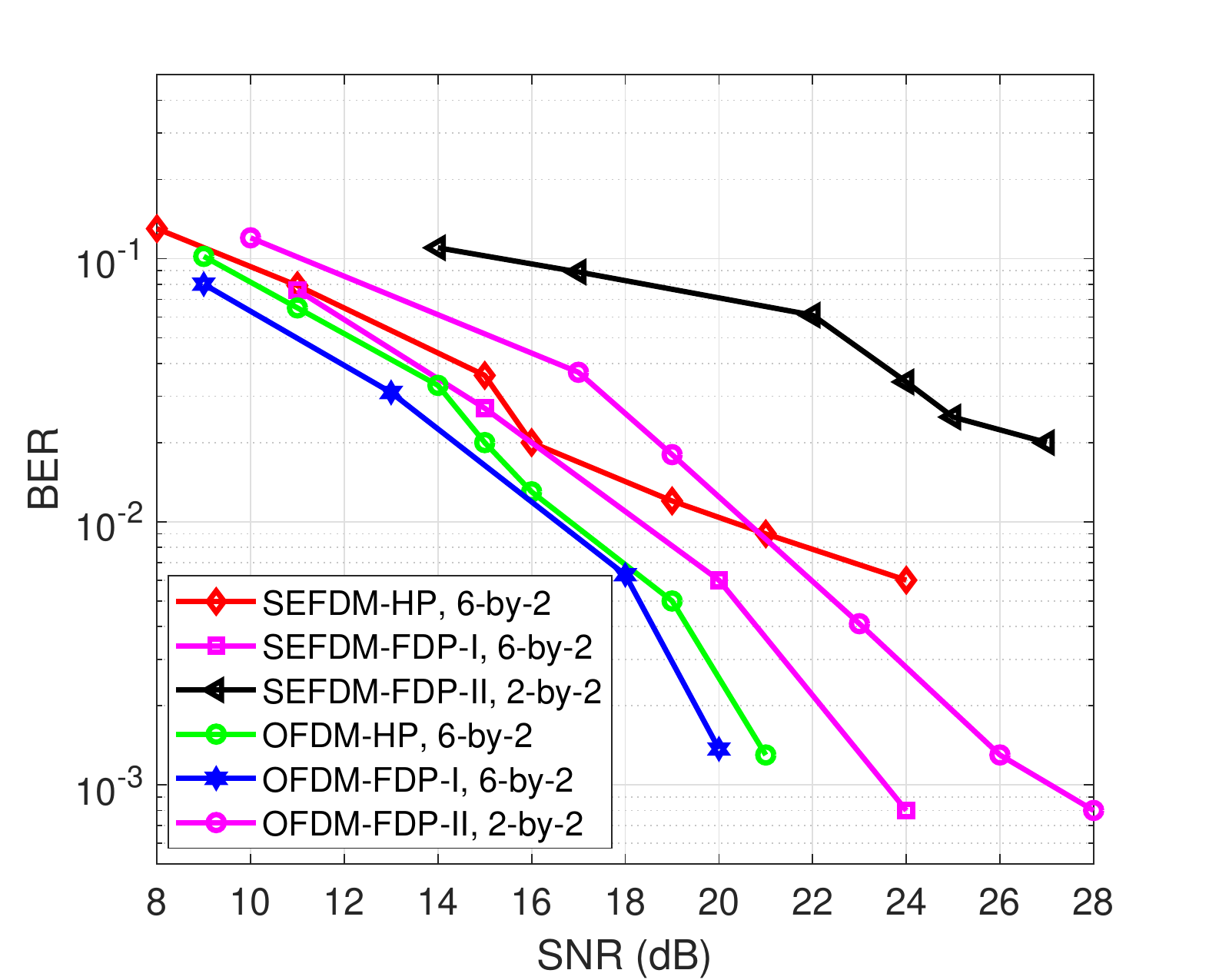}
\end{center}
\caption{Measured 16QAM BER. }
\label{Fig:Hybrid_precoding_SEFDM_BER_16QAM}
\end{figure}

\begin{figure}[t!]
\begin{center}
\includegraphics[scale=0.49]{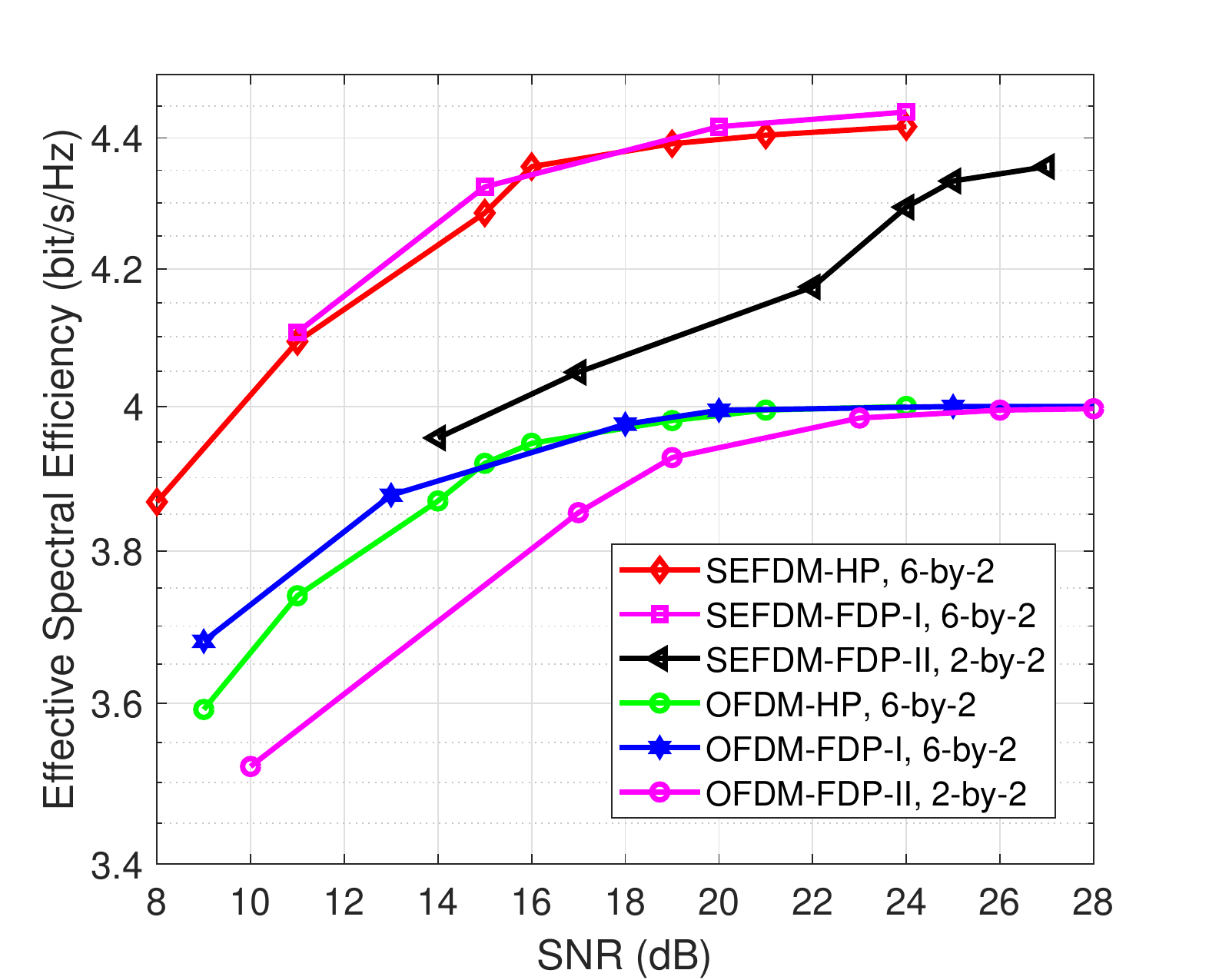}
\end{center}
\caption{Measured 16QAM spectral efficiency. }
\label{Fig:Hybrid_precoding_SEFDM_SE_16QAM}
\end{figure}

However, the 16QAM modulated systems show different results in Fig. \ref{Fig:Hybrid_precoding_SEFDM_BER_16QAM} and Fig. \ref{Fig:Hybrid_precoding_SEFDM_SE_16QAM}. Due to the high constellation density of 16QAM, two RF chains are not sufficient to precode spatially interfered OFDM signals resulting in performance degradation. With the hybrid precoding architecture, performance is improved with a narrow gap with the fully digital precoding OFDM system of six RF chains. The situation is even worse for SEFDM, in which \ac{ICI} would amplify the distortion. Therefore, for the SEFDM system of two RF chains, an error floor starts to appear. Even with the hybrid precoding architecture, its BER curve would not converge and result in an error floor as well. Only the fully digital precoding system of six RF chains can converge and approach the OFDM performance.

The divergence of BER performance determines the maximum spectral efficiency convergence speed. The OFDM system of six RF chains reaches the maximum spectral efficiency faster than other precoding OFDM systems. For the SEFDM system of two RF chains, since an error floor appears, its spectral efficiency would be unlikely to reach the peak value. In addition, there is a narrow gap between the six-RF-chain fully digital precoding SEFDM and the hybrid precoding SEFDM due to the BER performance loss between the two systems.

The testbeds for different signals are configured in various experimental arrangements of RF chains, omni-directional antennas and phase shifters, resulting in a set of performance parameters of \ac{EVM}, \ac{BER} and \ac{SE}. In principle, the fully digital FDP-I systems achieve the best performance given the higher number of RF chains employed. However, extra power consumption is required for extra hardware resources. Therefore, a more reasonable criterion to compare different systems is using energy efficiency. Work in \cite{hybrid_precoding_EE_2016} defined energy efficiency to be the ratio between spectral efficiency and  total power consumption. This experimental work follows the same definition of energy efficiency but with slight changes based on our practical testbed conditions.

In this experiment, highly integrated software defined radio devices USRP are used. It is difficult to isolate the USRP's power consumption parts, such as  RF chains, power amplifiers and other components. Therefore, in the total power consumption calculation, the power consumption $P_{usrp}$ for each USRP device is taken into account as a single figure. Since phase shifters are stand alone components in this experiment, its power consumption $P_{ps}$ is considered independently. The tailored energy efficiency calculation for our experiments is defined as
\begin{equation}
\eta=\frac{SE_e}{N_{usrp}{\cdotp}P_{usrp}+N_{ps}{\cdotp}P_{ps}},\label{eq:energy_efficiency}\end{equation}
where $N_{usrp}$ indicates the number of USRP devices employed in each experiment and $N_{ps}$ is the number of phase shifters.

\begin{figure}[t!]
\begin{center}
\includegraphics[scale=0.49]{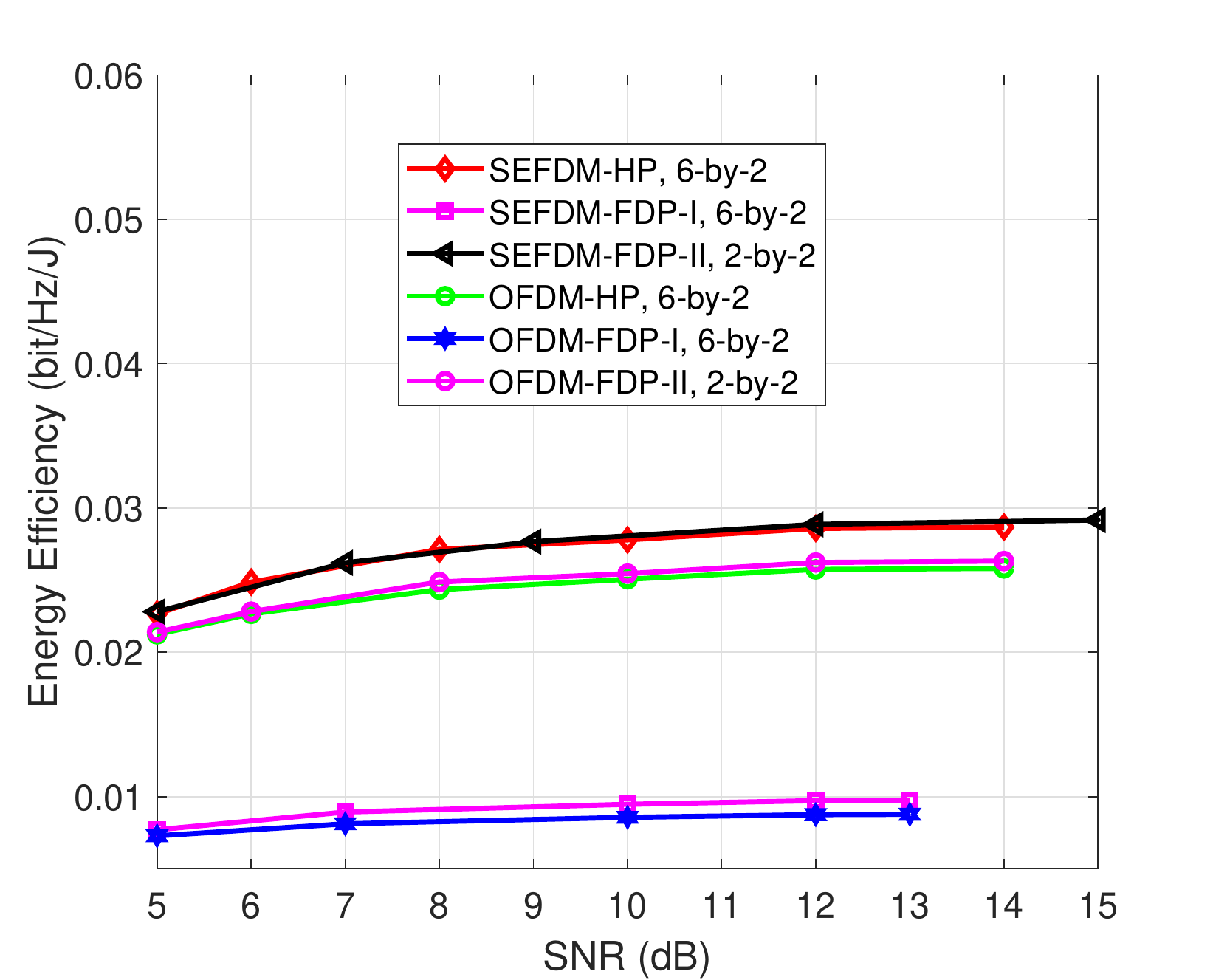}
\end{center}
\caption{Energy efficiency for 4QAM modulated signals. }
\label{Fig:Hybrid_precoding_SEFDM_EE_4QAM}
\end{figure}

\begin{figure}[t!]
\begin{center}
\includegraphics[scale=0.49]{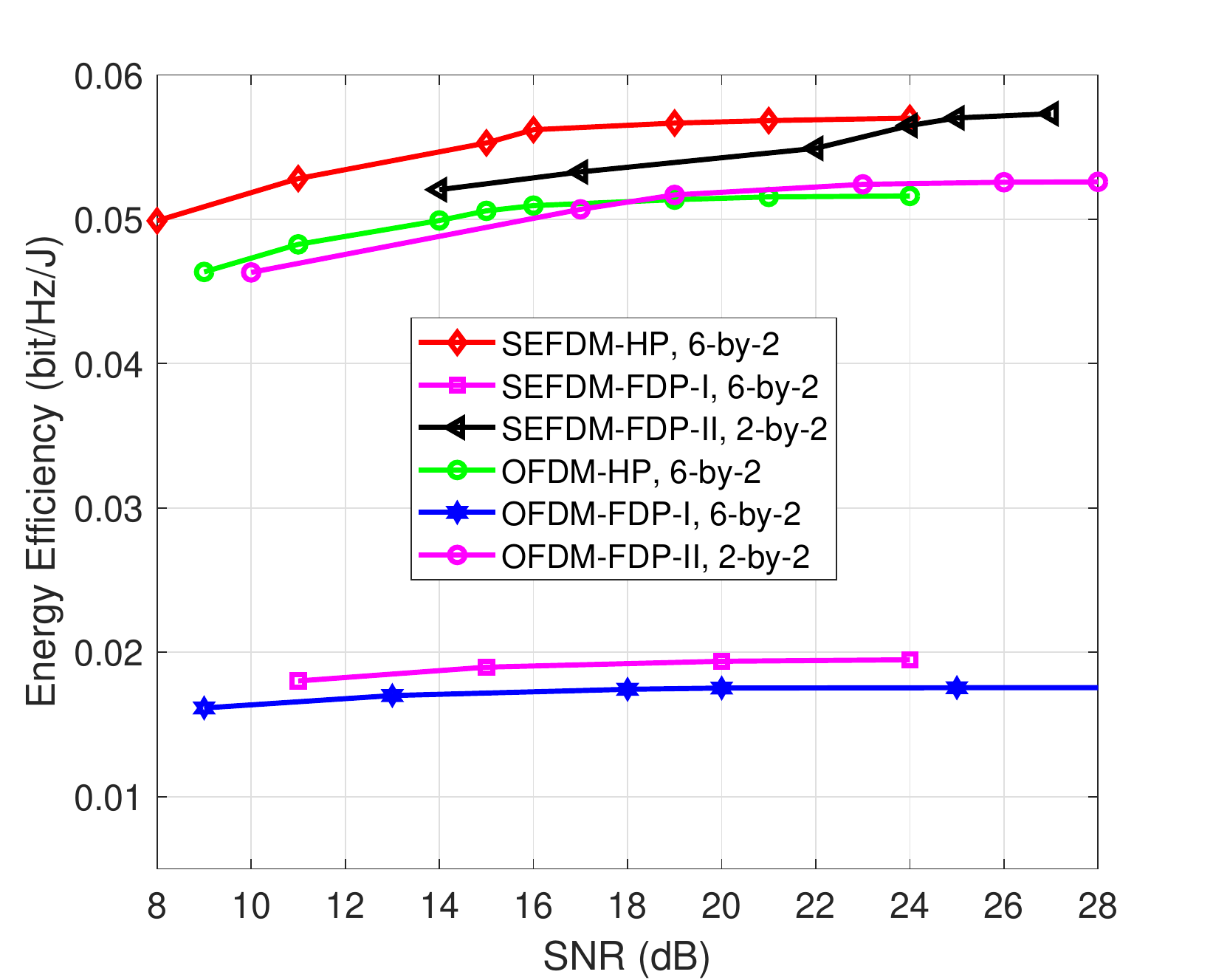}
\end{center}
\caption{Energy efficiency for 16QAM modulated signals. }
\label{Fig:Hybrid_precoding_SEFDM_EE_16QAM}
\end{figure}

According to the specifications of USRP-2953R \cite{USRP_RIO2953R} and Vaunix LPS-402 \cite{Vaunix_phase_shifter_ref}, the reasonable power consumptions for the USRP and the phase shifter are 38W and 250mW, respectively. Considering the effective spectral efficiency obtained in Fig. \ref{Fig:Hybrid_precoding_SEFDM_SE_4QAM} and Fig. \ref{Fig:Hybrid_precoding_SEFDM_SE_16QAM}, the energy efficiency levels for 4QAM and 16QAM modulated signals are calculated and illustrated in Fig. \ref{Fig:Hybrid_precoding_SEFDM_EE_4QAM} and Fig. \ref{Fig:Hybrid_precoding_SEFDM_EE_16QAM}, respectively. In principle, both results reveal that the hybrid precoding system outperforms its fully digital precoding counterpart in terms of energy efficiency, when considering the same antenna scales.

{Fig. \ref{Fig:Hybrid_precoding_SEFDM_EE_4QAM} reveals that the best performance achievable FDP-I architectures for both OFDM and SEFDM result in the worst energy efficiency performance. This is due to the use of extra USRP devices in the FDP-I architecture. The FDP-II architecture and the HP architecture have similar energy efficiency since the same number of USRP devices are employed for both systems and the dominant power consumption comes from the USRP. It should be noted that the FDP-II and HP architectures for SEFDM reach higher energy efficiency levels than that of OFDM because of the higher spectral efficiency achieved from Fig. \ref{Fig:Hybrid_precoding_SEFDM_SE_4QAM}.}

{For the 16QAM modulated signals in Fig. \ref{Fig:Hybrid_precoding_SEFDM_EE_16QAM}, the FDP-II architecture still maintains the worst energy efficiency for both OFDM and SEFDM signals. For the other two architectures, SEFDM apparently outperforms OFDM due to its higher achieved spectral efficiency. It should be noted that the hybrid precoding architecture starts to show its advantage over the FDP-II architecture at low and moderate SNR values. The HP and FDP-II will both approach the maximum energy efficiency with the increase of SNR. This indicates that the hybrid precoding architecture can reach the optimal energy efficiency faster than the fully digital FDP-II architecture.}

\section{Conclusion}

This work deals with an efficient hybrid precoding strategy for low-cost non-orthogonal IoT communications. Due to the self-created \ac{ICI} of the non-orthogonal IoT signal and the low-cost requirements for IoT devices, a practical three-stage precoding architecture is proposed and experimentally validated. In this work, we found that the precoding accuracy is related to the number of transmitter RF chains. Two RF chains are sufficient for pure digital precoding 4QAM-OFDM signals. However, with the increase to 16QAM, two RF chains are no longer efficient. For the non-orthogonal signal waveform \ac{SEFDM}, two RF chains are not large enough for both 4QAM and 16QAM modulation formats due to the self-created \ac{ICI} impact. This motivates us to use more RF chains to precode signals. However, the number of RF chains should not be too large due to the computational complexity and hardware costs. One alternative solution is to use hybrid precoding instead of the pure digital precoding. Experiments are conducted for three precoding architectures using different signal waveforms and modulation formats. Practical measurements reveal that the hybrid precoding can save hardware resources while maintaining similar performance, which indicates improved hardware efficiency and its flexibility for low-cost IoT applications.

\bibliographystyle{IEEEtran}
\bibliography{Tongyang_Ref}

\end{document}